\documentclass{article}
\pdfoutput=1
\usepackage{amssymb,graphicx,cancel}
\usepackage[margin=2cm]{geometry}
\usepackage{slashed}
\usepackage{color}
\usepackage[table]{xcolor}
\usepackage{graphicx}
\usepackage{tikz}
\usepackage{tikz-feynman}
\usepackage[utf8]{inputenc}
\usepackage{booktabs}
\usepackage{multirow}
\usepackage{amssymb}
\usepackage[export]{adjustbox}
\usepackage{floatrow}
\usepackage{blindtext}
\usepackage{amsmath}
\usepackage{booktabs}
\usepackage{multicol}
\usepackage{lipsum}
\usepackage{marvosym}
\usepackage{graphicx}
\usepackage{subfig}
\usepackage{multirow}
\usepackage{array}
\usepackage[normalem]{ulem}
\usepackage{tabularx}
\usepackage{jheppub}
\usepackage[utf8]{inputenc}
\usepackage{cleveref}

\newcommand{\eg}{\emph{e.g.~}}
\newcommand{\ie}{\emph{i.e.~}}

\newcommand{\Eq}[1]{Eq.~(\ref{#1})}
\newcommand{\Sec}[1]{Sec.~\ref{#1}}

\title{{\boldmath$Z'$} Mediated WIMPs: \\
Dead, Dying, or Soon to be Detected?}
\author[1,2]{Carlos Blanco,}
\author[3]{Miguel Escudero,}
\author[2,4,5]{Dan Hooper,}
\author[6]{and Samuel J.~Witte}

\affiliation[1]{University of Chicago, Department of Physics, Chicago, IL 60637, USA}
\affiliation[2]{University of Chicago, Kavli Institute for Cosmological Physics, Chicago, IL 60637, USA}
\affiliation[3]{King's College London, Department of Physics, Strand, London WC2R 2LS, UK}
\affiliation[4]{Fermi National Accelerator Laboratory, Theoretical Astrophysics Group, Batavia, IL 60510, USA}
\affiliation[5]{University of Chicago, Department of Astronomy and Astrophysics, Chicago, IL 60637, USA}
\affiliation[6]{Instituto de F\'{i}sica Corpuscular (CSIC-Universitat de Val\`{e}ncia), Paterna (Valencia), Spain}

\abstract{Although weakly interacting massive particles (WIMPs) have long been among the most studied and theoretically attractive classes of candidates for the dark matter of our universe, the lack of their detection in direct detection and collider experiments has begun to dampen enthusiasm for this paradigm. In this study, we set out to appraise the status of the WIMP paradigm, focusing on the case of dark matter candidates that interact with the Standard Model through a new gauge boson. After considering a wide range of $Z'$ mediated dark matter models, we quantitatively evaluate the fraction of the parameter space that has been excluded by existing experiments, and that is projected to fall within the reach of future direct detection experiments. Despite the existence of stringent constraints, we find that a sizable fraction of this parameter space remains viable. More specifically, if the dark matter is a Majorana fermion, we find that an order one fraction of the parameter space is in many cases untested by current experiments. Future direct detection experiments with sensitivity near the irreducible neutrino floor will be able to test a significant fraction of the currently viable parameter space, providing considerable motivation for the next generation of direct detection experiments.}

\emailAdd{carlosblanco2718@uchicago.ed}
\emailAdd{miguel.escudero@kcl.ac.uk}
\emailAdd{dhooper@fnal.gov}
\emailAdd{sam.witte@ific.uv.es}

\begin{document}

\preprint{FERMILAB-PUB-19-331-A, KCL-2019-61}

\maketitle

\section{Introduction}

Over the past several decades, the most popular and well-studied candidates for dark matter have been stable particles that were in equilibrium with the Standard Model (SM) bath in the early universe and that then froze-out to yield a thermal relic abundance in agreement with the measured cosmological dark matter density. In order for this process to result in an acceptable dark matter abundance, such particles were generally required to possess very roughly weak-scale masses and couplings to the SM. This result provided the foundation for what has become known as the WIMP paradigm. 

It has long been appreciated that if the dark matter consists of weakly interacting massive particles (WIMPs), it should be possible to detect these particles through their elastic scattering with nuclei, by observing their annihilation products, or by producing them in colliders (for recent reviews, see Refs.~\cite{Lin:2019uvt,Hooper:2018kfv,Cline:2018fuq}). With this goal in mind, large and highly sensitive underground detectors have been developed and deployed, resulting in very stringent limits on the dark matter's scattering cross section with nuclei~\cite{Aprile:2018dbl,Akerib:2016vxi,Cui:2017nnn,Aprile:2019dbj,Akerib:2017kat}. The Large Hadron Collider (LHC) has also begun to explore the electroweak-scale, but has not identified any evidence that dark matter particles are being produced in these collisions~\cite{Sirunyan:2018dub,Sirunyan:2018xlo,Sirunyan:2018fpy,Sirunyan:2018wcm,Sirunyan:2018gka,Sirunyan:2017leh,Aaboud:2018xdl,Aaboud:2017phn,Aaboud:2017rzf,Aaboud:2017bja,Aaboud:2017yqz}. Lastly, while the Galactic Center gamma-ray excess~\cite{Daylan:2014rsa,Calore:2014xka,Goodenough:2009gk,Hooper:2010mq,Hooper:2011ti,Abazajian:2012pn,Hooper:2013rwa,TheFermi-LAT:2015kwa,Leane:2019xiy} and the cosmic-ray antiproton excess~\cite{Cholis:2019ejx,Cuoco:2019kuu,Cuoco:2016eej,Cui:2016ppb} are each suggestive of originating from dark matter annihilation, no consensus has emerged regarding the interpretation of this data. These results have motivated many scientists working on the problem of dark matter to consider alternatives to the WIMP paradigm~\cite{Bertone:2018xtm}, elevating the degree of interest being directed towards candidates such as axions~\cite{Agrawal:2017eqm,Agrawal:2017cmd,Blinov:2019rhb,Arias:2012az,Marsh:2015xka,Graham:2015ouw,Irastorza:2018dyq}, as well as scenarios in which the dark matter is part of a hidden sector~\cite{Pospelov:2007mp,ArkaniHamed:2008qn,Abdullah:2014lla,Berlin:2014pya,Martin:2014sxa,Hooper:2012cw,Cline:2014dwa,Berlin:2016vnh,Berlin:2016gtr,Dror:2016rxc,Dror:2017gjq,Feng:2009mn,Cheung:2010gj,Chu:2011be,Escudero:2016ksa,Zurek:2013wia,Escudero:2017yia,Elor:2018twp}.

At this point in time, it is not entirely clear how one should view the status of the WIMP paradigm. On the one hand, it is certainly the case that many once attractive dark matter candidates have been excluded by the null results of direct detection experiments and by searches for new physics at the LHC~\cite{Escudero:2016gzx,Arcadi:2017kky,Balazs:2017ple}. It is also true, however, that many varieties of WIMPs remain entirely viable~\cite{Gelmini:2016emn,Plehn:2017fdg,Roszkowski:2017nbc,Kowalska:2018toh}. How one thinks about the relative weighting of these scenarios impacts how we should devote our experimental and theoretical resources. With so much at stake, we would ideally attempt to make a systematic and thorough assessment of the current status of the WIMP paradigm. Given the vast diversity of possible WIMP models that one could consider, however, a truly exhaustive study would be an enormous and practically intractable undertaking. With such considerations in mind, we have chosen to focus more narrowly in this study on the case of dark matter particles that annihilate through couplings to a new vector gauge boson, $Z'$~\cite{Agashe:2007jb,Agashe:2004bm,Agashe:2004ci,Lee:2007mt,Buckley:2011mm,Buckley:2011vs,Belanger:2007dx,Hur:2007ur,DelleRose:2017ukx,DelleRose:2017uas,Lebedev:2014bba,Arcadi:2013qia,Fairbairn:2016iuf,Dudas:2013sia,Chu:2013jja,Mambrini:2011dw,Dudas:2009uq,Hooper:2014fda}. New broken $U(1)$ gauge symmetries and the $Z'$ bosons that accompany them are found within many well-motivated extensions of the SM~\cite{Langacker:2008yv}, including many Grand Unified Theories (GUTs)~\cite{London:1986dk,Hewett:1988xc} and string-inspired models~\cite{Braun:2005bw,Cleaver:1998gc,Coriano:2007ba,Faraggi:1990ita,Giedt:2000bi,Lebedev:2007hv,Anastasopoulos:2006cz,Faraggi:1991mu,Cvetic:2001nr}, as well as within the context of dynamical symmetry breaking scenarios~\cite{Hill:2002ap,Chivukula:2003wj,Chivukula:2002ry}, models with extra spatial dimensions~\cite{Agashe:2003zs,Agashe:2007ki,Carena:2003fx,Hewett:2002fe}, and many other popular extensions of the SM~\cite{ArkaniHamed:2001nc,Cvetic:1997ky,Langacker:1999hs,ArkaniHamed:2001is,ArkaniHamed:2002qx,Han:2003wu,Perelstein:2005ka}. Within this relatively simple subset of WIMP models, we will consider scenarios in which the dark matter candidate is either a Majorana or Dirac fermion, and $Z'$ bosons that possess a wide range of couplings and other characteristics. This collection of well-motivated models can lead to a wide range of phenomenological consequences, with detection prospects that vary from easily testable, to extremely elusive.

The remainder of this study is structured as follows.  In Sec.~\ref{sec:models}, we describe the range of $Z'$ mediated dark matter models that we will consider in this study. We then describe in Sec.~\ref{pheno} the current and projected constraints that we apply to this class of models. In Sec.~\ref{results} we present our main results. After discussing some caveats and other theoretical considerations in Sec.~\ref{caveats}, we attempt in Sec.~\ref{bayesian} to quantitatively evaluate the status of $Z'$ mediated WIMPs. To this end, we perform a Bayesian analysis, calculating for each given model (and for three choices of priors) the fraction of the parameter space that has been ruled out by existing experiments, as well as the fraction that is projected to fall within the reach of future direct detection experiments. Although the current constraints do exclude a significant fraction of the $Z'$ mediated dark matter parameter space, a sizable proportion remains viable (in the case that the dark matter is a Majorana fermion). The prospects for future direct detection experiments are quite encouraging; we project that experiments with sensitivity near the neutrino floor will be able to test a significant fraction of the currently viable parameter space. We discuss and summarize our results in Sec.~\ref{conclusion}.

\section{$Z'$ Mediated Dark Matter Models}\label{sec:models}

In this section, we describe the range of $Z'$ mediated dark matter models considered in this study. In order to ensure maximum generality, we have taken a simplified models approach, in which we describe the masses and couplings of the dark matter and $Z'$ without necessarily specifying the full particle content of the underlying theory. Although one might ideally like to consider models that are UV complete and fully gauge invariant~\cite{FileviezPerez:2010gw,Duerr:2013dza,Duerr:2014wra,Ismail:2016tod,Casas:2019edt,Ellis:2017tkh,Arcadi:2013qia,Ellis:2018xal,Caron:2018yzp,Madge:2018gfl,FileviezPerez:2018jmr,FileviezPerez:2019jju,ElHedri:2018cdm,Escudero:2018fwn,Foldenauer:2018zrz,Bagnaschi:2019djj,Das:2019pua}, this comes at the cost of significantly increasing the dimensionality of the parameter space. Here, we will consider models that respect the symmetries of the SM and maintain tree-level gauge invariance, but do not explicitly require the cancellation of gauge anomalies. Within the context of such models, we assume that loop-level gauge invariance is achieved through the presence of additional unspecified particles, which do not play a significant role in the dark matter phenomenology under consideration. For additional discussion, see Sec.~\ref{caveats}.

\subsection{Dirac Dark Matter}
\label{diracmodel}

The simplest realization containing a Dirac dark matter candidate, $\chi$, arises when the $Z^\prime$ acquires its mass through the Stuckelberg mechanism (see, for example, Ref~\cite{Alves:2015mua}). Here, the Lagrangian is extended by the following (neglecting the dark matter kinetic term):
\begin{equation}
\mathcal{L} \supset -  \sum_i g^\prime \, q_{i}\, Z^\prime_\mu\,\bar{f_i} \gamma^\mu f_i + m_\chi \bar{\chi}\chi - \frac{\epsilon}{4} \, F^{\mu\nu} \, F^{\prime}_{\mu\nu} - \frac{1}{4} \, F^{\prime \, \mu\nu}\, F^{\prime}_{\mu\nu},
\end{equation}
where the sum is performed over all SM fermions as well as the dark matter candidate. The quantities $g^\prime$, $F^{\prime}_{\mu\nu}$ and $q_{i}$ are the gauge coupling, field strength tensor, and charge assignments of the $U(1)'$, respectively. The kinetic mixing between the $U(1)'$ and $U(1)_Y$ is quantified by $\epsilon$, which we take to be zero at tree-level (but is induced through loops, as described in Sec.~\ref{epsilonloop}). For simplicity, we we will often refer to the interactions of the $Z'$ in terms of its effective universal coupling to SM fermions, $g_{SM} \equiv q_{i} g^\prime$ (where $i$ includes all SM fermions that are charged under the $U(1)'$), and it coupling to the dark matter, $g_{\chi} \equiv q_\chi g^\prime$.

\subsection{Majorana Dark Matter} 

In the case of dark matter in the form of a Majorana fermion, one cannot simply exploit the Stuckelberg mechanism, as simplified $Z^\prime$ models with non-zero axial couplings naturally violate unitarity at high energies~\cite{Kahlhoefer:2015bea}. This problem can be circumvented, however, if one instead generates the necessary masses through the spontaneous breaking of the $U(1)'$ symmetry by a new SM singlet scalar, $\phi$, which we take here to be complex and charged under the new $U(1)^\prime$ with $q_\phi = 2 q_\chi$. Specifically, we will assume that the Lagrangian in the unbroken phase contains the following terms:
\begin{eqnarray}
\label{lagmaj}
\mathcal{L} &\supset& - \sum_i g^\prime \, q_{i}\, \, Z^\prime_\mu\,\bar{f}_i \gamma^\mu f_i - \frac{1}{2} g^\prime q_\chi Z^\prime_\mu \bar{\chi} \gamma^\mu \gamma_5 \chi -  \frac{\lambda_{\chi}}{\sqrt{2}} \left( \phi \, \overline{\chi} \chi^c  +h.c.\right)  \\ 
&+& (D_\mu \phi)^\dagger D^\mu \phi +  \mu_\phi^2 \phi^\dagger \phi - \lambda_\phi (\phi^\dagger \phi)^2 - \lambda_{H\phi}H^\dagger H \phi^\dagger \phi - \frac{\epsilon}{4}\, F^{\mu\nu} \, F^{\prime}_{\mu\nu} - \frac{1}{4} \, F^{\prime \, \mu\nu}\, F^{\prime}_{\mu\nu},  \nonumber
\end{eqnarray}
where $\lambda_\chi$ is a Yukawa coupling, $\mu_\phi^2$ and $\lambda_\phi$ are parameters in the scalar potential, and $\lambda_{H\phi}$ is the scalar-Higgs mixing. We again take $\epsilon$ to be zero at tree-level, and additionally assume that the scalar-Higgs mixing vanishes ($\lambda_{H\phi} = 0$). Spontaneous symmetry breaking causes the scalar to develop a vacuum expectation value, $v^\prime$. In the unitary gauge, one can rewrite the field as $\phi = \frac{1}{\sqrt{2}} (v^\prime + \rho)$, where $\rho$ is a CP-even scalar field. Minimization of the scalar potential yields $\mu_\phi^2 = \lambda_\phi v^{\prime \, 2}$. By substituting $\phi = \frac{1}{\sqrt{2}} (v^\prime + \rho)$ into Eq.~\ref{lagmaj}, together with $D_\mu \phi = \partial_\mu \phi -  i  g^\prime  2 q_\chi Z^\prime_\mu \phi$, one finds that the resulting Lagrangian contains:
\begin{align}
\mathcal{L} &\supset - \sum_i g^\prime \, q_{i} \, Z^\prime_\mu\,\bar{f}_i \gamma^\mu f_i - g^\prime \frac{q_\chi}{2} Z^\prime_\mu \bar{\chi} \gamma^\mu \gamma_5 \chi - \frac{\lambda_\chi}{2}(v^\prime + \rho) \bar{\chi}\chi  \\
&  + \frac{1}{2}\partial_\mu \rho \partial^\mu \rho +  2 \, g^{\prime \, 2} q_\chi^2 Z^\prime_\mu Z^\prime {}^\mu \left( v^{\prime \, 2} + 2 v^\prime \rho +\rho^2\right) - \frac{1}{4} \lambda_\phi (\rho+v^\prime)^2 \left(\rho^2+2 \rho v^\prime-v^{\prime \, 2}\right).  \nonumber
\end{align}
In the broken phase, the mass of the dark matter, new gauge boson, and real scalar can be expressed as follows: $m_\chi = \lambda_\chi v^\prime$, $m_{Z^\prime} = 2 q_\chi g^\prime v^\prime$ and $m_{\rho}^2 = 2 \lambda_\phi v^{\prime \, 2}$. Substituting in these mass parameters, one arrives at:
\begin{align}
\mathcal{L} \supset &- \sum_i g^\prime \, q_{i} \, Z^\prime_\mu\,\bar{f}_i \gamma^\mu f_i - g^\prime \frac{q_\chi}{2} Z^\prime_\mu \bar{\chi} \gamma^\mu \gamma_5 \chi - \frac{m_\chi}{2}\left(1 + \frac{\rho}{v^\prime}\right) \bar{\chi}\chi  \\
&  + \frac{1}{2}\partial_\mu \rho \partial^\mu \rho +  \frac{m_{Z^\prime}^2}{2} Z^\prime_\mu Z^\prime {}^\mu \left( 1 + \frac{\rho}{v^\prime}\right)^2 - \frac{m_\rho^2 }{8 v^{\prime \, 2}} \rho^2 (\rho+2 v^\prime )^2.   \nonumber
\end{align}
In order to minimize its impact of the resulting phenomenology, we will take the mass of the scalar to be equal to the maximum value consistent with unitarity, $m_{\rho} = \sqrt{\pi} \, m_{Z'}/g_{\chi}$ (see Appendix~\ref{sec:pwu}).

\subsection{Loop-Induced Kinetic Mixing}
\label{epsilonloop}

Kinetic mixing between the $U(1)'$ and $U(1)_Y$ can shift the mass and couplings of the $Z$ from their predicted value~\cite{Peskin:1991sw}, and thus precision electroweak measurements can be used to constrain the value of $\epsilon$~\cite{Langacker:1991pg,Umeda:1998nq,Babu:1997st}. With this in mind, we assume throughout this study that $\epsilon$ vanishes at tree level, but is generated at loop level, yielding the following~\cite{Holdom:1985ag,Bell:2014tta}:
\begin{eqnarray}
\label{epsilon}
\epsilon& \sim &\frac{g_Y g'}{12\pi^2}\sum_i Y_i \, q_i\ln \left(\frac{\Lambda^2}{m^2_{f_i}}\right),
\end{eqnarray}
where $g_Y$ is the SM gauge coupling, $Y_i$ is the hypercharge of fermion $i$, and $\Lambda = m_Z^\prime/\sqrt{g_\chi g_f}$ is the effective cutoff scale.

In addition to any tree-level couplings that may exist, kinetic mixing will induce an effective coupling of the $Z'$ to SM fermions: $ \mathcal{L} \in - g_Y \cos \theta_W \, \epsilon \, \bar{f} \gamma_\mu {f} Z'_{\mu}$, where $\epsilon \simeq g_{\rm SM} \, g_Y \cos \theta_W /4\pi^2 \sim 10^{-2} g_{\rm SM}$. These loop-induced couplings will play an important role in determining many of the constraints presented in this study and are included in all of the relevant calculations presented here.



\section{Dark Matter Phenomenology}
\label{pheno}

In this section, we describe our analysis of the $Z'$ mediated dark matter models presented in the previous section. In order to make this problem more tractable, we will limit our analysis to the following sets of $U(1)'$ charge assignments:

\begin{itemize}
\item{Coupling to lepton number, with $q_l=1$ for all SM leptons.}
\item{Coupling only to first-generation leptons, with $q_e=q_{\nu_e}=1$.}
\item{Coupling only to third-generation leptons, with $q_{\tau}=q_{\nu_{\tau}}=1$.}
\item{Coupling to baryon number, with $q_q=1/3$ for all SM quarks.}
\item{Coupling only to first-generation quarks, with $q_u=q_d=1/3$.}
\item{Coupling only to third-generation quarks, with $q_t=q_b=1/3$.}
\end{itemize}

We have chosen this selection of charge assignments in order to cover a diverse and representative range of phenomenological possibilities. For example, models without tree-level couplings to SM quarks (\ie ``leptophilic'' models) are generally less constrained by direct detection. Furthermore, models with couplings only to first or third generation fermions can lead to very different annihilation cross sections and scattering rates with nuclei (for theoretical motivation for models with couplings only to third generation fermions, see Refs.~\cite{Andrianov:1998hx,Hill:1994hp,Chivukula:2002ry,delAguila:1986iw,Belanger:2007dx}). While one could easily construct a $U(1)'$ model with charge assignment that do not fall within any of the above listed examples, the phenomenology of such a model would in most cases map closely onto one or more of the models considered here.

For each choice of charge assignments, we explore a 4-dimensional parameter space in terms of $m_\chi$, $m_{Z^\prime}$, $g_{\rm SM}$ and $g_{\chi}$. In each case, we consider four discrete values for $g_{\chi} / g_{\rm SM}$, equal to $10^{-2}$, $10^{-1}$, 1 and 10. Although these scenarios should perhaps not all be considered to be equally well-motivated, the choices of these ratios provides a broad perspective and allows one to observe how the various constraints are impacted by the choice of $g_{\chi} / g_{\rm SM}$. In general, scenarios featuring small values of $g_{\chi} / g_{\rm SM}$ are more strongly constrained, while larger values make the dark matter and $Z'$ increasingly secluded from the SM, in the limiting case constituting a hidden sector model~\cite{Pospelov:2007mp,ArkaniHamed:2008qn,Abdullah:2014lla,Berlin:2014pya,Martin:2014sxa,Hooper:2012cw,Berlin:2016vnh,Berlin:2016gtr,Dror:2016rxc,Dror:2017gjq,Feng:2009mn,Cheung:2010gj,Chu:2011be,Escudero:2016ksa,Zurek:2013wia,Elor:2018twp}. For each choice of $m_{\chi}$, $m_{Z'}$, $g_{\chi} / g_{\rm SM}$ and charge assignments, we select the value of $g_{\rm SM} \, g_{\chi}$ such that the thermal relic abundance is equal to the measured cosmological dark matter density, $\Omega_{\chi} h^2= 0.12$~\cite{Aghanim:2018eyx}, as calculated using the publicly available program \texttt{micrOMEGAs} (version 5.0.4)~\cite{Belanger:2013oya}. We then assess whether a given point in parameter space is consistent with the constraints from direct detection, indirect detection, measurements of the cosmic microwave background (CMB), and a variety of collider, fixed target and neutrino experiments.

\subsection{Model Requirements}
\label{apriori}

Throughout this study, we will remain largely agnostic regarding the masses and couplings of the dark matter candidate and the $Z'$. There are,  however, a number of model independent requirements that we can impose on these parameters. Firstly, we require that partial wave unitarity is respected, as described in Appendix~\ref{sec:pwu}. We also require each coupling in the theory to be smaller than $\sqrt{4\pi}$, in order to maintain perturbativity. And lastly, we require that the width of the $Z'$ does not exceed 10\% of its mass, $\Gamma_{Z'} < 0.1 \,m_{Z'}$.\footnote{Dark matter annihilation cross sections are computed in \texttt{micrOMEGAs}~\cite{Belanger:2013oya} under the assumption that all particles involved in the annihilation processes have a narrow width. Therefore, for consistency, we require the width of the $Z'$ not to exceed 10\% of its mass, $\Gamma_{Z'} < 0.1 \, m_{Z'}$. Furthermore, since $\Gamma_{Z'} \sim g^2/(8\pi) \,m_{Z'}$, regions of parameter space in which $\Gamma_{Z'} > 0.1 \, m_{Z'}$ correspond to $g \gtrsim 0.45 \times \sqrt{4\pi}$, only marginally consistent with the requirement of perturbativity.}

\subsection{Constraints from Cosmology}
\label{cosmology}

Measurements of the temperature anisotropies in the cosmic microwave background (CMB) and of the primordial light nuclei abundances enable us to place important constraints on the parameter space within this class of models. In particular, throughout this study we will consider only parameter space with $m_{\chi}, m_{Z'} \gtrsim 10$ MeV, in order to avoid conflict with the successful predictions of Big Bang Nucleosynthesis (BBN)~\cite{Boehm:2013jpa,Nollett:2014lwa,Escudero:2018mvt}. 

The annihilation of dark matter particles in the era leading up to and after recombination can have an observable impact on the CMB. More specifically, the annihilation products can produce large numbers of ionizing photons, which increase the fraction of free electrons in the universe. This has a direct impact on the integrated optical depth as observed by Planck, which directly constrains the annihilation power at the 95\% CL, defined as~\cite{Aghanim:2018eyx}:
\begin{align}
p_{\rm ann} \equiv f_{\text{eff}}\frac{\langle \sigma v\rangle}{m_{\chi}} < 3.4 \times 10^{-28} \;\text{cm}^3/\text{s}/\text{GeV},
\end{align}
where the effective efficiency factor, $f_{\text{eff}}$, is the fraction of the annihilation power that is transferred into the intergalactic medium during the relevant range of redshifts~\cite{ade2016planck,Aghanim:2018eyx}. For a given model, we calculate $f_{\text{eff}}$ by integrating the $e^{\pm}$ and gamma-ray annihilation spectra as calculated by \texttt{micrOMEGAs}~\cite{Belanger:2013oya} (utilizing PYTHIA~\cite{Sjostrand:2014zea}) over the precalculated $f_{\mathrm{eff}}^{e^\pm ,\gamma}$ curves provided in Ref~\cite{Slatyer:2015jla}:
\begin{align}
f_{\mathrm{eff}}\left(m_{\chi}\right)=\frac{1}{2 m_{\chi}} \int_{0}^{m_{\chi}}\left(f_{\mathrm{eff}}^{e} \frac{d N_e}{d E_{e}}+f_{\mathrm{eff}}^{\gamma} \frac{d N_{\gamma}}{d E_{\gamma}}\right) E d E.
\end{align}
This procedure yields a bound that generally rules out $s$-wave annihilating dark matter with $m_\chi \lesssim 10-20$ GeV (see, however, Ref.~\cite{Leane:2018kjk}).


\subsection{Direct Detection}

Searches for the elastic (or inelastic) scattering of dark matter particles with nuclei have provided some of the most powerful constraints on WIMPs. In recent years, experiments utilizing a target of liquid xenon (including XENON1T~\cite{Aprile:2018dbl,Aprile:2019dbj}, LUX~\cite{Akerib:2016vxi,Akerib:2017kat}, and PandaX-II~\cite{Cui:2017nnn}) have placed the most stringent constraints on such interactions across much of the relevant parameter space.  

For each model under consideration, we compute the leading order scattering cross section and compare it with the 90\% CL upper limit obtained from the aforementioned experiments. In cases in which this interaction occurs at tree level, the cross section is computed using \texttt{micrOMEGAs}. In models in which the $Z'$ does not couple to quarks, however, scattering with nuclei only occurs through loop-induced interactions arising from kinetic mixing. Such scattering is dominated by the heaviest charged lepton that couples to the $Z'$, and leads to the following cross section for the cases of Dirac and Majorana dark matter, respectively~\cite{Bell:2014tta}:
\begin{align}
\sigma_{\rm Dirac} &=\frac{\mu_{N}^{2}}{9 \pi}\left[\frac{\alpha_{\rm EM} Z}{\pi \Lambda^{2}} \log \left(\frac{m_{\ell}^{2}}{\Lambda^{2}}\right)\right]^{2},  \label{DDleptophilicVV} \\
\sigma_{\rm Majorana} &=\frac{\mu_{N}^{2} v_{\chi}^{2}}{9 \pi}\left(1+\frac{\mu_{N}^{2}}{2 m_{N}^{2}}\right)\left[\frac{\alpha_{\rm EM} Z}{\pi \Lambda^{2}} \log \left(\frac{m_{\ell}^{2}}{\Lambda^{2}}\right)\right]^{2}, \nonumber
\label{DDleptophilicAV}
\end{align}
where $\mu_N \equiv m_N m_\chi/(m_N + m_\chi)$ is the reduced mass of the nucleus-dark matter system, $Z$ is the charge of the nucleus, $v_\chi \sim 10^{-3}\, c$ is the velocity of the dark matter, and $\Lambda = m_{Z'}/\sqrt{g_{\chi} g_l}$. 


\subsection{Indirect Detection}

Indirect searches include efforts to detect the gamma rays, antiprotons, positrons, neutrinos and other particles that are produced in the annihilations (or decays) of dark matter particles. In this study, we apply constraints as derived from gamma-ray observations of the Milky Way's dwarf spheroidal galaxies by the Fermi telescope~\cite{Fermi-LAT:2016uux} and measurements of the cosmic-ray $e^{\pm}$ spectrum by AMS-02~\cite{Aguilar:2014mma,Accardo:2014lma}.

To apply these constraints, we use \texttt{micrOMEGAs} to calculate the spectrum of gamma rays, electrons and positrons that are produced per annihilation in a given model and compare with the 95\% CL upper limits obtained from Fermi-LAT and AMS-02. There is a high degree of complementarity between these measurements, as Fermi is most sensitive to annihilations that produce quarks or tau leptons, while AMS-02 yields its strongest constraints in the case of annihilations to muons or electrons. In models in which $m_\chi > m_{Z^\prime}$ and $g_{\chi} \gg g_{\rm SM}$, the $t$-channel annihilation into a pair of on-shell $Z^\prime$ bosons can be the dominant annihilation channel. In this case, the boosted decay of the $Z'$ annihilation products leads to a rather smooth $e^{\pm}$ spectrum, without the distinctive spectral features that are present in models featuring direct annihilation to $e^+ e^-$ or $\mu^+ \mu^-$~\cite{Elor:2015bho,Elor:2015tva}. In this case, the AMS-02 constraints are significantly weakened, and are not included here.


\subsection{Collider, Fixed Target and Neutrino Experiments}

The results of accelerator experiments have been used to place stringent constraints on the mass and couplings of a $Z'$. For relatively heavy $Z'$ bosons, some of the strongest limits come from searches for dijet resonances at experiments including CMS, ATLAS, CDF and UA2~\cite{Aaboud:2017yvp,Khachatryan:2016ecr,Sirunyan:2016iap,Sirunyan:2017nvi,Aad:2014aqa,Khachatryan:2015sja,Dobrescu:2013coa}. Such searches provide particularly stringent constraints on the couplings of a $Z^\prime$ to quarks. Searches at the LHC for dilepton resonances also broadly constrain models in which the $Z'$ couples more strongly to charged leptons~\cite{Aaboud:2017buh,Sirunyan:2018exx}. We apply these constraints rescaling the bounds derived in Ref.~\cite{Escudero:2018fwn} by the appropriate model-dependent production factor and branching ratios. We also apply constraints derived from the measurement of LEP, which strongly limit the couplings of a $Z'$ to electrons. For $m_{Z'} \gtrsim 200$ GeV, LEP provides a limit of $g_{e} < (m_{Z'}/\,7 \,{\rm  TeV})$~\cite{Carena:2004xs}. 

For the case of a lighter $Z'$, a wide range of constraints have been derived from the results of collider and beam dump experiments, including BaBar, NA48/2, LHCb, KLOE, NA64, as well as electron and proton beam dumps~\cite{Merkel:2014avp,Abrahamyan:2011gv,Ablikim:2017aab,Lees:2014xha,Bergsma:1985qz,Bjorken:1988as,Riordan:1987aw,Bross:1989mp,Adrian:2018scb,Konaka:1986cb,Anastasi:2015qla,Anastasi:2016ktq,Anastasi:2018azp,Babusci:2012cr,Babusci:2014sta,Aaij:2017rft,Batley:2015lha,Banerjee:2018vgk,Astier:2001ck,Blumlein:1990ay,Blumlein:1991xh,Davier:1989wz,Bernardi:1985ny,Lees:2017lec,Fox:2011fx,Banerjee:2016tad,Banerjee:2017hhz,NA64:2019imj}. We apply this collection of constraints to the specific $Z'$ models considered here using the DarkCast software~\cite{Ilten:2018crw}. In addition, we also apply the following constraints on the couplings to leptons as derived from Borexino data~\cite{Bellini:2011rx}: $g_e < 5 \times 10^{-3} \, (m_{Z'}/{\rm GeV})$~\cite{Harnik:2012ni} and $(g_{\mu,\tau} \, \epsilon \, g_Y \,\cos \theta_W)^{1/2} < 5\times 10^{-3} \,(m_{Z'}/{\rm GeV}) $~\cite{Kamada:2015era}. These constraints account for the effects of kinetic mixing as well as the fact that roughly 33\% of the solar neutrino flux is of each flavor.

Note that the constraints obtained from the LHC, DarkCast, and Borexino do not appear in the summary plots of the Dirac dark matter candidates; this is not to suggest that they don't apply or exist, but rather this is a reflection of the fact that the parameter spaced probed by these sources is excluded by a combination of other experiments.

All of the limits from collider, fixed target, and neutrino experiments are quoted at the 95\% CL.

\subsection{Reach of Future Direct Detection Experiments}
\label{neutrinofloor}

Direct detection experiments will ultimately encounter an irreducible background arising from the coherent scattering of the ambient neutrino background~\cite{Billard:2013qya,Ruppin:2014bra}. This background of neutrinos is produced from various sources, including nuclear reactions in the Sun~\cite{Robertson:2012ib,Serenelli:2011py}, interactions of cosmic rays in the atmosphere~\cite{Gaisser:2002jj}, galactic supernovae~\cite{Horiuchi:2008jz,Beacom:2010kk}\footnote{Typically only the diffuse isotropic supernovae background is included in calculations of neutrino background. Should a local star core collapse, however, pre-~\cite{Raj:2019wpy} and post-supernovae~\cite{Chakraborty:2013zua,XMASS:2016cmy,Lang:2016zhv,Kozynets:2018dfo,Khaitan:2018wnf} neutrinos may also contribute. Since these signals are rare and strongly time-dependent, we neglect these contributions in what follows.}, nuclear fission reactors~\cite{reines1953detection,reines1960detection,Hayes:2016qnu}, and decays of radioactive elements in the Earth~\cite{Araki:2005qa,Bellini:2010hy}. The signature produced by the coherent scattering of these neutrinos is remarkably similar to what is naively expected for dark matter, and will consequently inhibit the ability of direct detection experiments to probe new parameter space. It is important to emphasize, however, that the so-called ``neutrino floor'' is not entirely impregnable, as the spectrum of recoils produced by coherent neutrino scattering is, in general, not entirely degenerate with the recoil spectrum predicted from dark matter.
%
Experiments can thus, in principle, attempt to subtract this background~\cite{Billard:2013qya,Ruppin:2014bra,Davis:2014ama,Gelmini:2018ogy}. In this regime, however, constraints on the cross section would be expected to scale more slowly with exposure. It is thus unclear as to whether there will exist sufficient motivation to build experiments that are capable of significantly cutting into the neutrino floor. For the purposes of this work, we will define the final stage of direct detection as the maximally optimistic realizations of currently proposed experiments. We describe these experiments below, and summarize their properties in Table~\ref{tab:experiments}.

{\bf Argon G3:} The most futuristic proposal made by the DarkSide collaboration is the construction of a 300 tonne (200 tonne fiducial volume) argon time projection chamber that would operate for up to five years (this proposal is being referred to ``ARGO'')~\cite{zuzel2017darkside,Aalseth:2017fik}. We adopt two different operating thresholds for this experiment, one consistent with the high-energy regime outlined in the ARGO proposal, the other being a low-mass search consistent with the recent analysis performed by the Darkside-50 collaboration~\cite{Agnes:2018ves}. 

{\bf Xenon G3:} The XENON collaboration has proposed an experiment, referred to as ``DARWIN'', intended to extend in sensitivity all the way to the atmospheric neutrino background. The current proposal assumes a 40 tonne fiducial volume of liquid xenon operating for five years~\cite{Aalbers:2016jon}. The current design documents list a threshold of $\sim 5$ keV, adopted in order to avoid the solar neutrino background at low energies (current xenon experiments achieve an absolute threshold closer $1.1$ keV, albeit with limited efficiency)~\cite{Akerib:2016vxi,Aprile:2017iyp,Cui:2017nnn,Aprile:2018dbl}. We optimistically adopt a threshold of 1 keV, although we emphasize that the constraints we derive at low masses from other experiments are more stringent, and thus our results are not strongly sensitive to this choice.

{\bf Fluorine G3: } The conceptual design for the construction of PICO-500, a $\sim$\,$500$ kg fiducial volume bubble chamber, has recently been approved by SNOLAB~\cite{picoweb,pico_pres}. In our calculations we adopt a 6 keV threshold and a 2 tonne-year exposure, although we emphasize that our final result is only slightly sensitive to these choices, as the spin-dependent bound derived from the Xenon G3 experiment is typically stronger for models in which the $Z'$ couples equally to neutrons and protons. 

{\bf Germanium G3: } The CDMS collaboration has published estimated sensitivity curves for their next generation experiment, SuperCDMS SNOLAB. This experiment is expected to begin operation in 2020, and is not expected to reach the neutrino floor. Thus, we consider an advanced version of this experiment comprised of the germanium high-voltage detectors, for which the current threshold is $\sim$\,$40$ eV~\cite{Agnese:2016cpb}, and an exposure of 5 tonne-years.

\begin{table}[tb]
	\begin{center}
		\vspace*{0cm}
		\begin{tabular}[c]{l|ccc} \hline\hline
			Target   & Exposure & $E_{\rm th}$&  $E_{\rm max}$ \\ & [tonne-year] & [keV] & [keV] \\
 \hline
             Argon G3 (ARGO)  & $10^3$  & 10.0 &  150 \\
             Argon G3 (ARGO S2)  & $10^3$  & 0.6 &  10 \\ 
             Xenon G3  (DARWIN) & 200  & 1.0 &  150 \\		          
             Fluorine G3 (PICO-500) & 2  & 6.0 &  150 \\		          
		    Germanium G3 (SuperCDMS$+$)  & 5  & 0.04 &  50 \\		          
\hline\hline
		\end{tabular}
		\caption{\label{tab:experiments} Configurations adopted for our projection of neutrino-floor direct detection experiments. The projected sensitivities of these experiments are shown in Fig.~\ref{nuFloor}.}
		\vspace*{-0.5cm}
	\end{center}
\end{table}

\medskip

\begin{figure}[h]
\centering
\includegraphics[width=.495\textwidth]{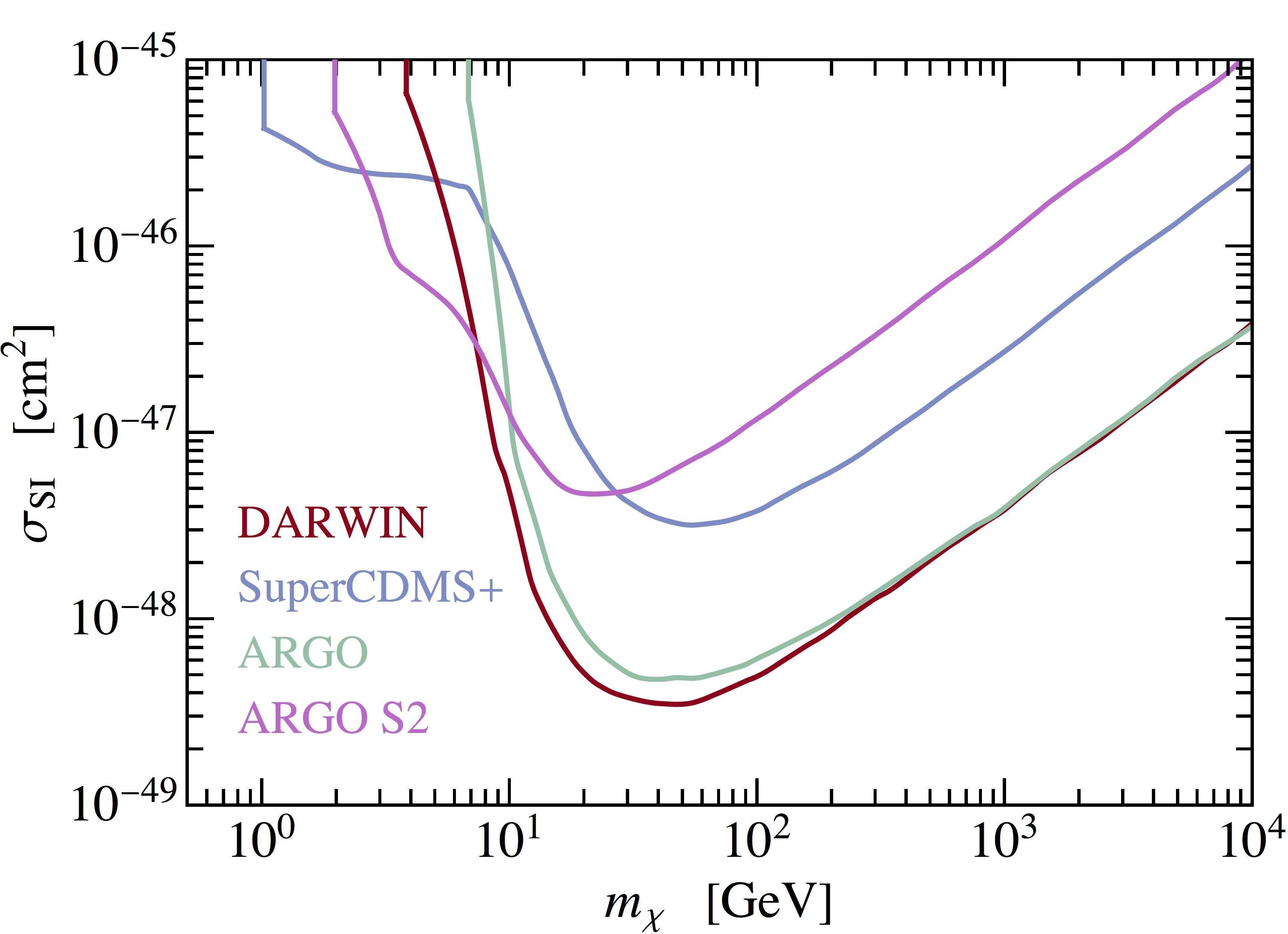}
\includegraphics[width=.495\textwidth]{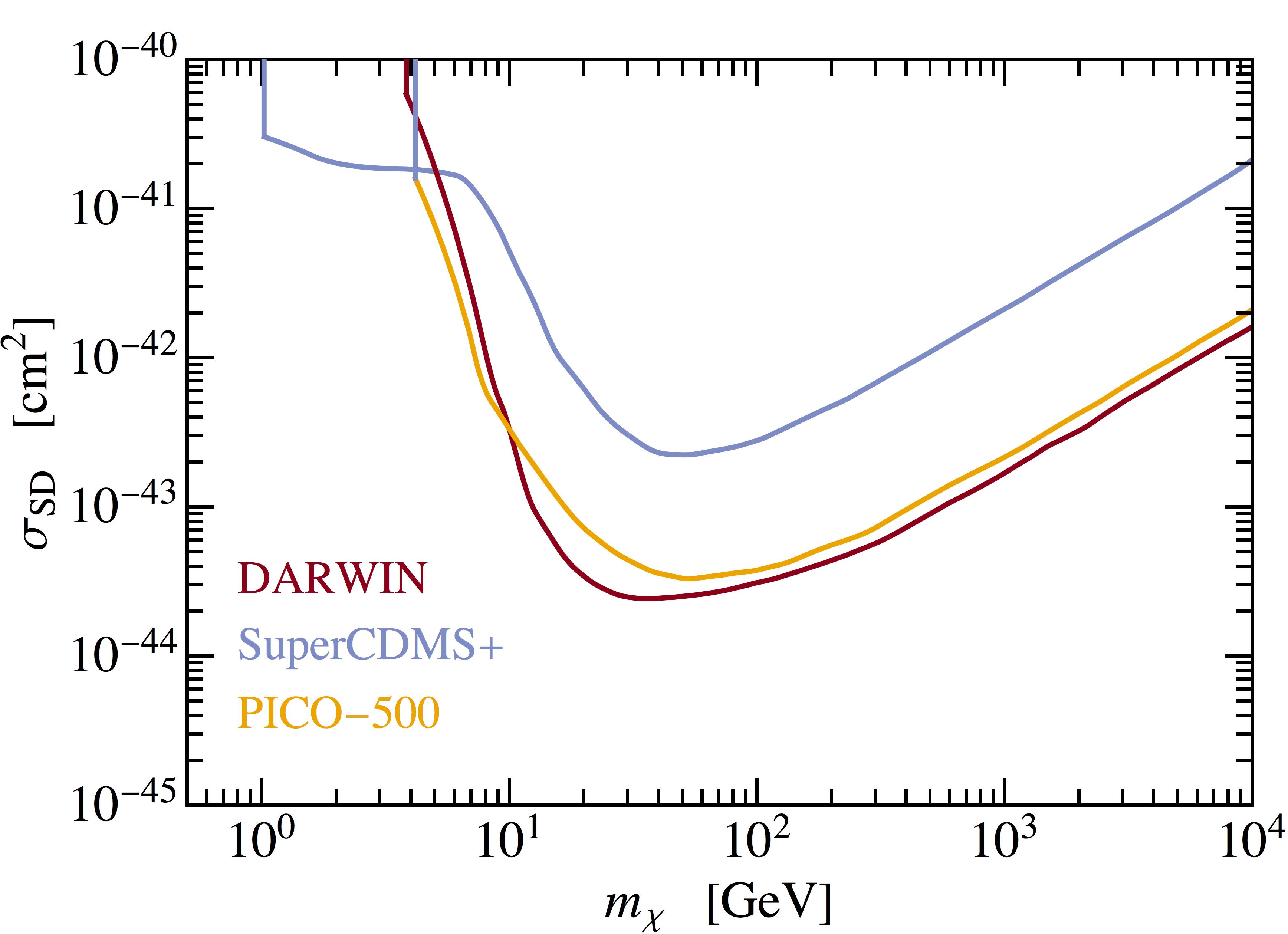}
\caption{The projected neutrino-floor sensitivity for several direct detection experiments, for the case of equal couplings to protons and neutrons. Constraints are shown for spin-independent (left) and spin-dependent (right) scattering.}
\label{nuFloor}
\end{figure}

In order to project the sensitivity of the above described experiments, we simulate neutrino events for each of the experimental realizations using the neutrino fluxes provided in~\cite{Gelmini:2018ogy}\footnote{It is worth mentioning that the flux of reactor~\cite{Gelmini:2018ogy}, geological~\cite{huang2013reference,Gelmini:2018gqa}, and atmospheric neutrinos~\cite{Honda:2006qj} depend, in principle, on the geographic location of the experiment. Reactor and geological neutrinos are expected to be a subdominant background, thus we adopt the fluxes appropriate for the SNOLAB mine with the understanding that this choice will have a minimal impact on our results. For atmospheric neutrinos, we adopt the so-called FLUKA flux~\cite{Battistoni:2005pd} tabulated at Kamioka Mine as this is the only location for which low energy atmospheric fluxes have been computed. It is worth noting, however, that the differences between various cites can differ by a factor of $\sim$\,2-3 at low energies where the atmospheric neutrino flux is relevant for direct detection experiments~\cite{Honda:2006qj}. }, and derive $90\%$ upper limits on the direct detection cross sections using an extended likelihood function. This procedure is repeated $10^3$ times, each time producing new realizations of the neutrino data. For each dark matter mass and interaction, we identify the minimum cross section constrained by at least $90\%$ of the realizations. These projected bounds represent our effective neutrino floor, and are shown in Fig.~\ref{nuFloor} for the case of equal couplings to protons and neutrons.  

\section{Results}
\label{results}

\subsection{Dirac Dark Matter}


In this section, we consider the case of dark matter in the form of a Dirac fermion, as described in Sec.~\ref{diracmodel}. For each of the charge assignments described in Sec.~\ref{pheno}, we scan over $m_{\chi}$ and $m_{Z'}$ and consider four discrete choices of $g_{\chi}/g_{\rm SM}$. At each point in this parameter space, we set the product of these couplings, $g_{\chi}g_{\rm SM}$ such that the desired thermal relic abundance is obtained, $\Omega_{\chi} h^2 \simeq 0.12$.

\subsubsection{Couplings to Quarks}

We begin with the case of a $Z'$ that couples equally to all SM quarks. Such a scenario could arise, for example, in a model in which baryon number is gauged~\cite{Carone:1994aa,FileviezPerez:2010gw,Duerr:2013dza,Perez:2014qfa,Duerr:2014wra,Duerr:2013lka}. In such models, the dark matter annihilates to quark-antiquark pairs without velocity-suppression, leading to indirect detection constraints that are sensitive to masses up to $m_{\chi} \sim 60-70$ GeV. Even more significantly, this model features unsuppressed spin-independent scattering with nuclei, resulting in extremely stringent direct detection constraints.

\begin{figure}[H]
\centering
\textbf{Dirac Dark Matter, Couplings to all Quarks}\par\medskip
\centerline{\includegraphics[scale=0.68]{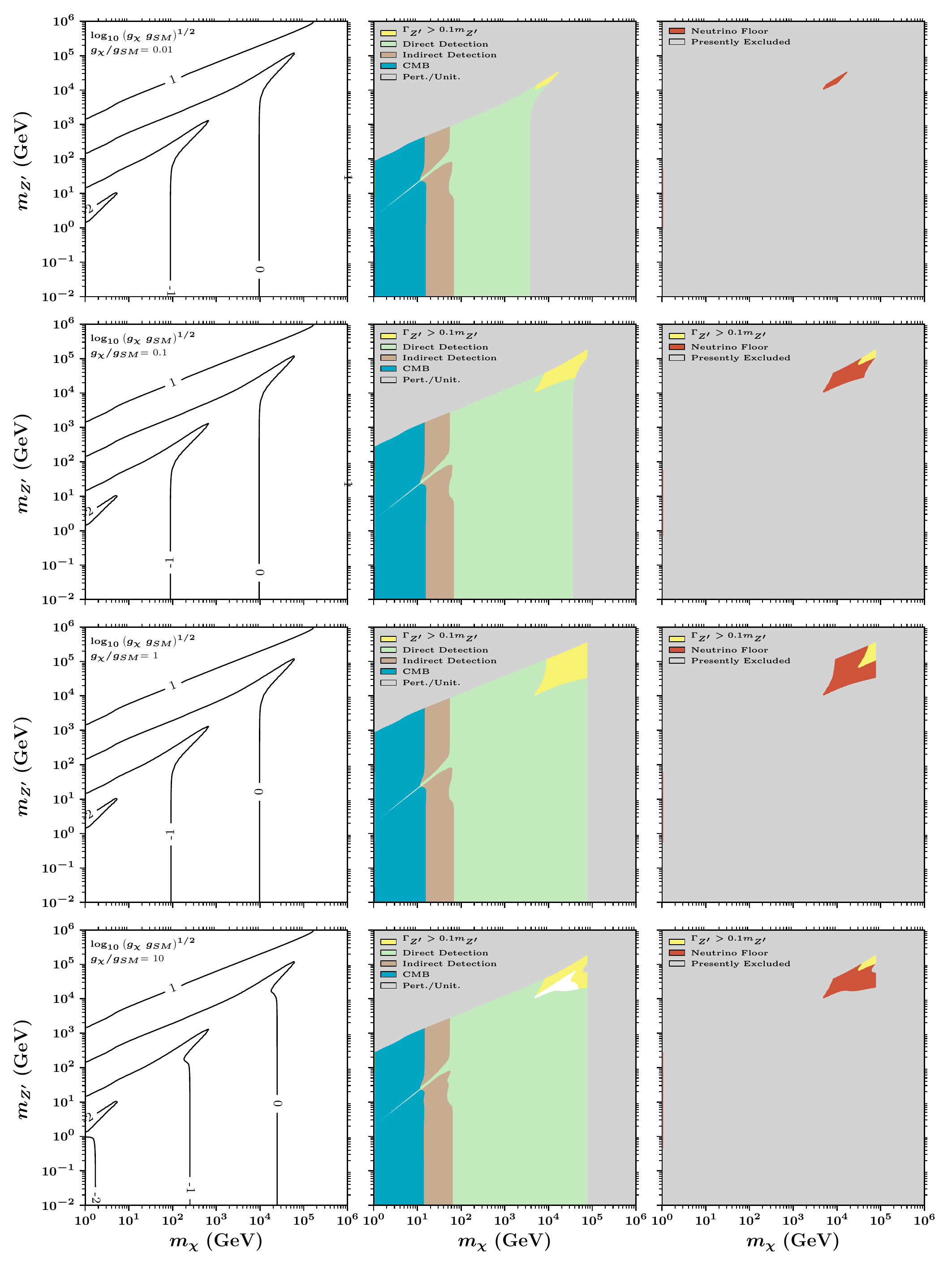}}
\caption{Constraints and prospects for detecting Dirac dark matter that is coupled to a $Z'$ with couplings to all SM quarks. In the left frames, we plot the values of $g_{\chi} g_{\rm SM}$ that yield $\Omega_{\chi} h^2 \simeq 0.12$. In the center and right frames, we show the current and projected constraints on this class of models, respectively. In each row, a different value of $g_{\chi}/g_{\rm SM}$ has been adopted. The combined constraints from the cosmic microwave background, direct detection and indirect detection rule out the overwhelming majority of the parameter space shown. Direct detection limits are shown at the 90\% CL, while all other experiments are shown at the 95\% CL.}
\label{Leptophobic_Dirac_All}
\end{figure}

\begin{figure}[h]
\centering
\includegraphics[width=.495\textwidth]{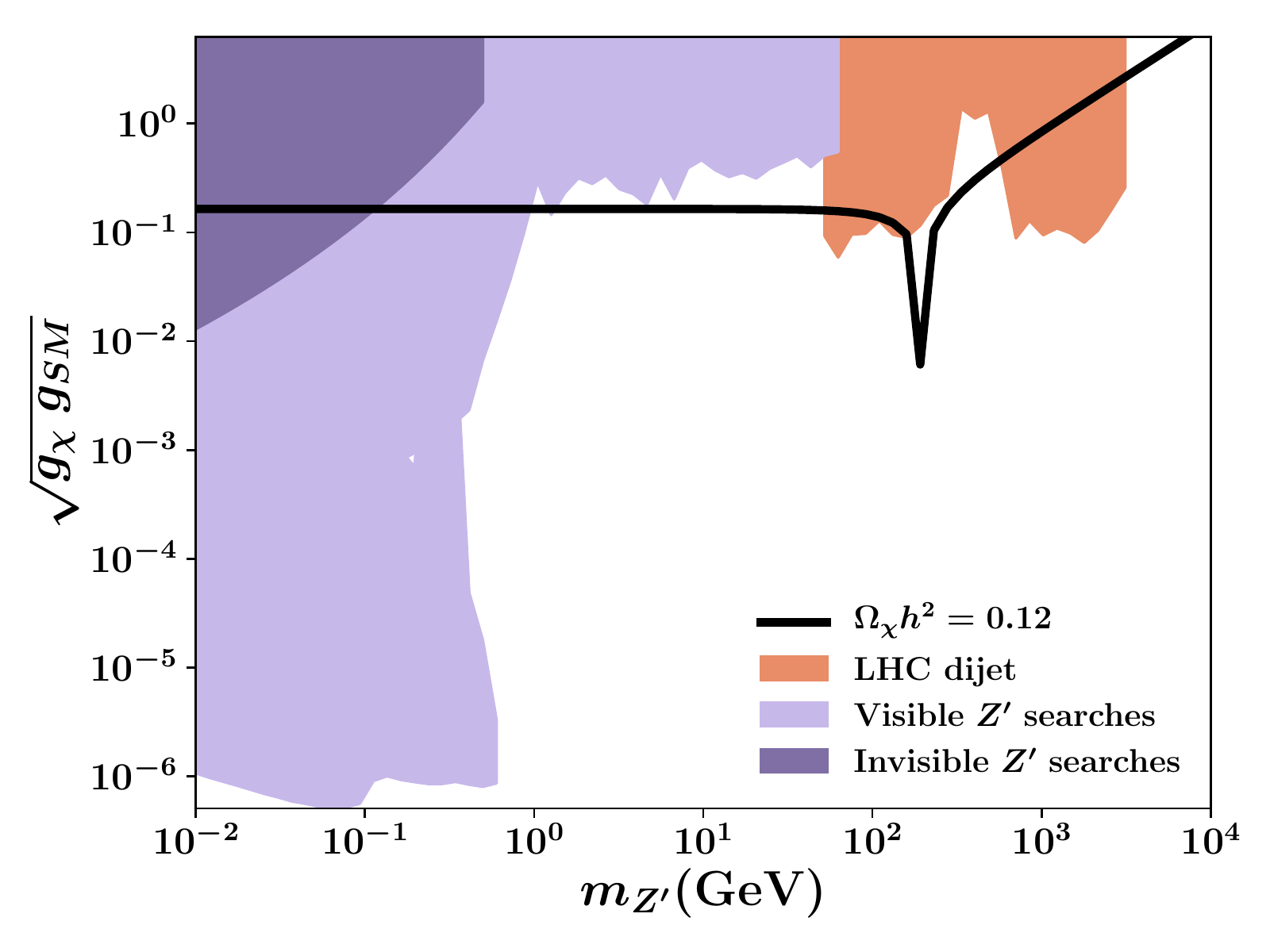}
\includegraphics[width=.495\textwidth]{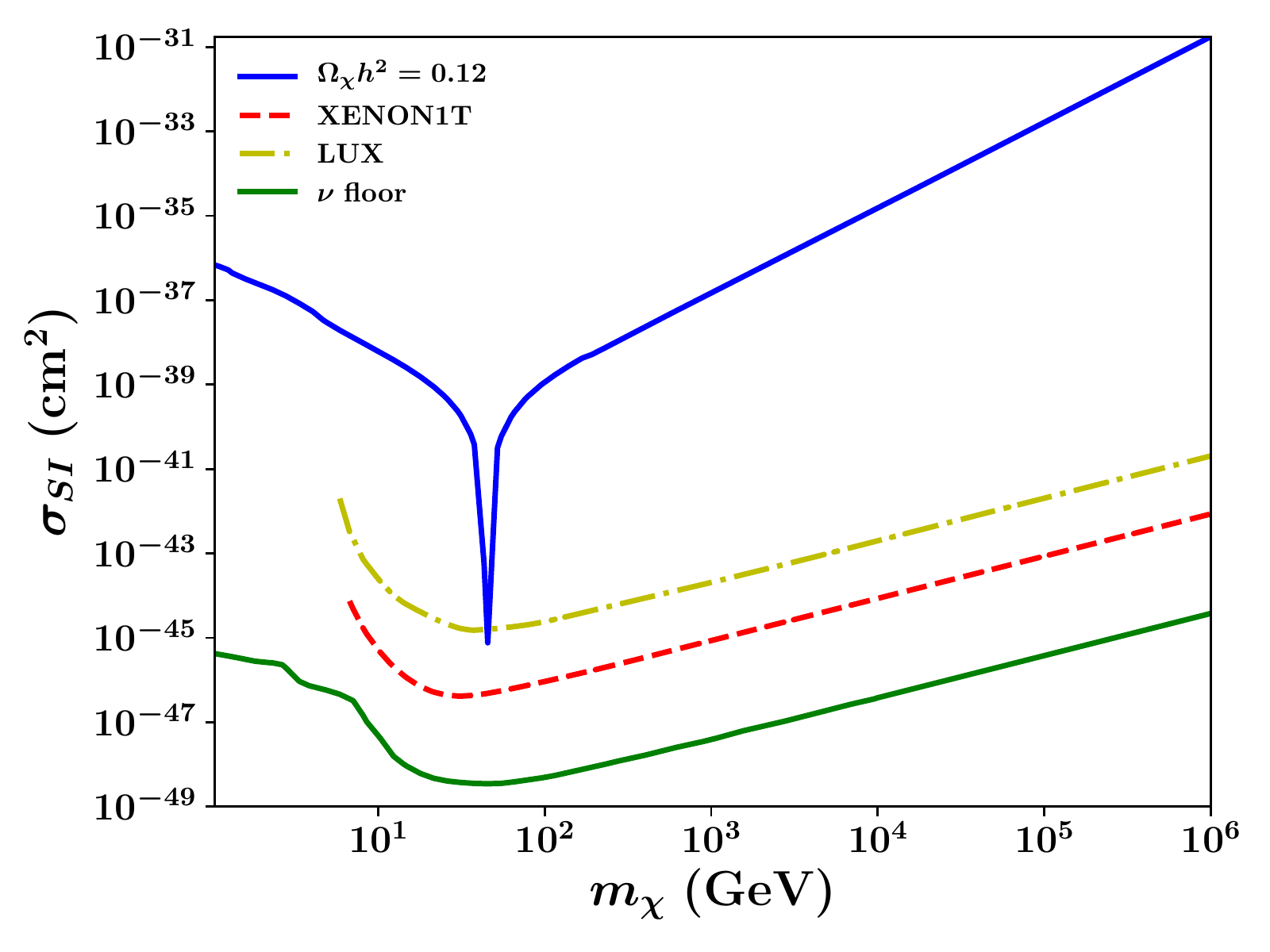}
\caption{Left Frame: Constraints on Dirac dark matter that is coupled to a $Z'$ with couplings to all SM quarks, for the case of $m_{\chi}=100$ GeV and $g_{\chi}/g_{\rm SM}=1$. Searches for light $Z'$ bosons~\cite{Ilten:2018crw} exclude $Z'$ masses below $\sim$\,1 GeV, while dijet searches at ATLAS and CMS exclude some regions of parameter space with larger values of $m_{Z'}$. Right frame: The spin-independent elastic scattering cross section with nuclei in the same model, for the case of $m_{Z'}=100$ GeV and $g_{\chi}/g_{\rm SM}=1$. Current direct detection experiments exclude the range of models shown for all values of $m_{\chi}$ above the threshold for XENON1T, LUX and PandaX-II. Direct detection limits are shown at the 90\% CL, while all other experiments are shown at the 95\% CL. }
\label{slices41}
\end{figure}

In Fig.~\ref{Leptophobic_Dirac_All} we summarize the current and projected constraints on this class of models. In each of the left frames, we plot contours of constant $\log_{10} \sqrt{g_{\chi} g_{\rm SM}}$ that yield the desired thermal relic abundance, $\Omega_{\chi} h^2 \simeq 0.12$. For these choices for the product of the couplings, we then plot in the center frames the current constraints on this model, as described in Sec.~\ref{pheno}. We discard those regions labeled ``Pert./Unit.'' on the grounds that they are not consistent with the requirements of perturbativity and unitary, as well as those labeled $\Gamma_{Z'} >0.1 \, m_{Z'}$. The combined constraints from the cosmic microwave background, direct detection and indirect detection rule out the overwhelming majority of the parameter space of this model. The only scenarios that are not currently excluded are those in which the dark matter mass lies very near the $Z'$ resonance ($m_{Z'} \simeq 2 m_{\chi}$) with large values of $m_{\chi}$ and $g_{\chi}/g_{\rm SM}$. Although we have chosen to plot the results of this model only above $m_{\chi} > 1$ GeV, the constraints provided by measurements of the CMB exclude all dark matter masses below this value. Also, although collider and fixed target experiments constrain parts of the parameter space shown, those regions are also excluded by current direct detection experiments, and thus do not appear in this figure. 

In the right frames of Fig.~\ref{Leptophobic_Dirac_All}, we illustrate the regions of the remaining parameter space that are projected to fall within the reach of future neutrino-floor direct detection experiments, as described in Sec.~\ref{neutrinofloor}. Such experiments are expected to fully explore the remaining parameter space in this case.

The constraints on this class of models are further illustrated in Fig.~\ref{slices41}, where we plot the results across specific slices of parameter space. For the case of $m_{\chi} = 100$ GeV and $g_{\chi}/g_{\rm SM} =1$, searches for light $Z'$ bosons (as characterized using DarkCast~\cite{Ilten:2018crw}) exclude $Z'$ masses below $\sim$\,1 GeV, while dijet searches at ATLAS and CMS exclude regions of parameter space with larger values of $m_{Z'}$. The most stringent constraints, however, are provided by direct detection experiments, which strongly exclude the range of models shown for all values of $m_{\chi}$ above the threshold for XENON1T, LUX and PandaX-II.

\begin{figure}[H]
\centering
\textbf{Dirac Dark Matter, Couplings to First Generation Quarks}\par\medskip
\centerline{\includegraphics[scale=0.68]{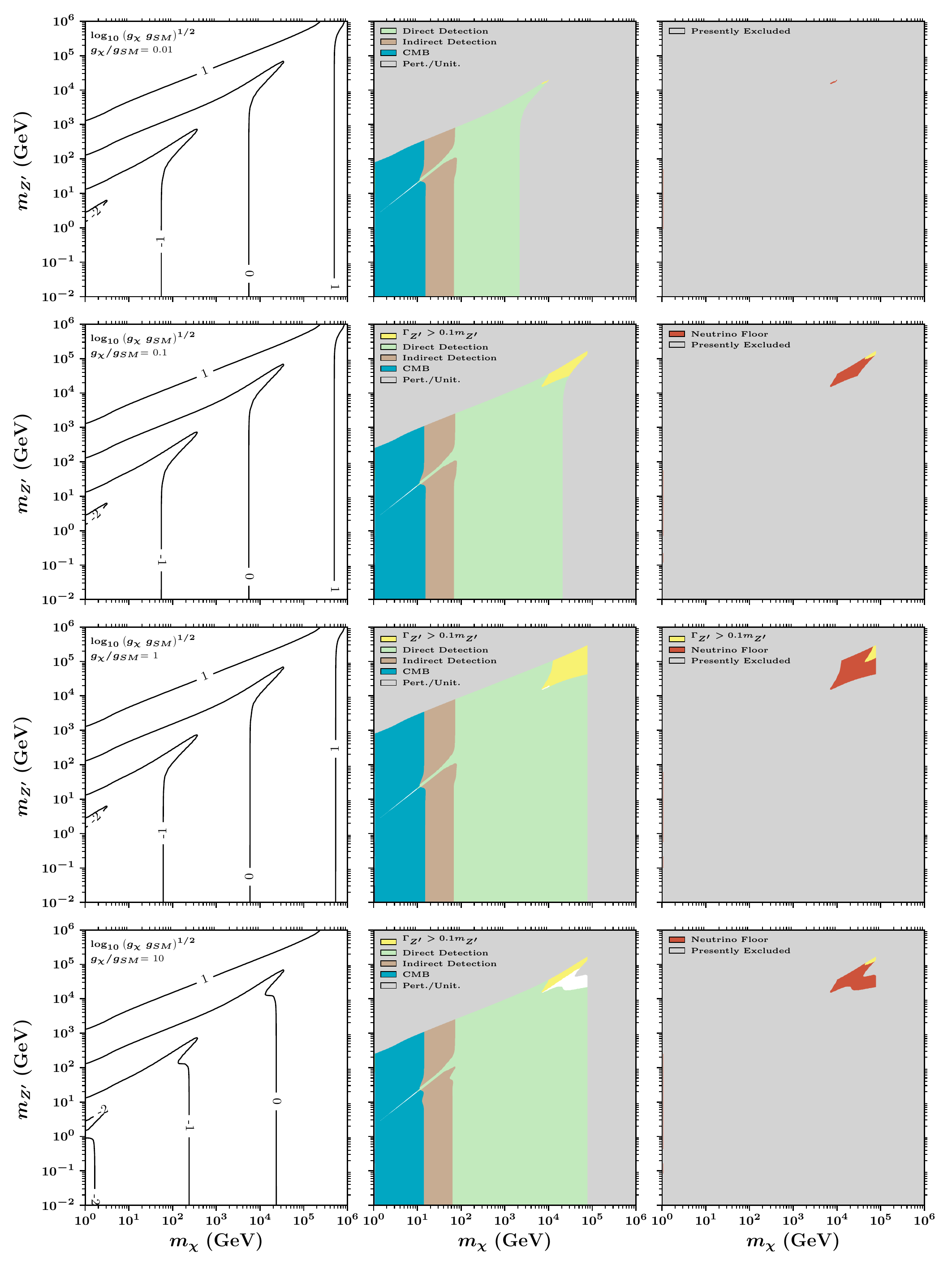}}
\caption{As in Fig.~\ref{Leptophobic_Dirac_All}, but for the case of Dirac dark matter that is coupled to a $Z'$ with couplings only to first generation quarks.}
\label{Leptophobic_Dirac_1st}
\end{figure}

\begin{figure}[H]
\centering
\textbf{Dirac Dark Matter, Couplings to Third Generation Quarks}\par\medskip
\centerline{\includegraphics[scale=0.68]{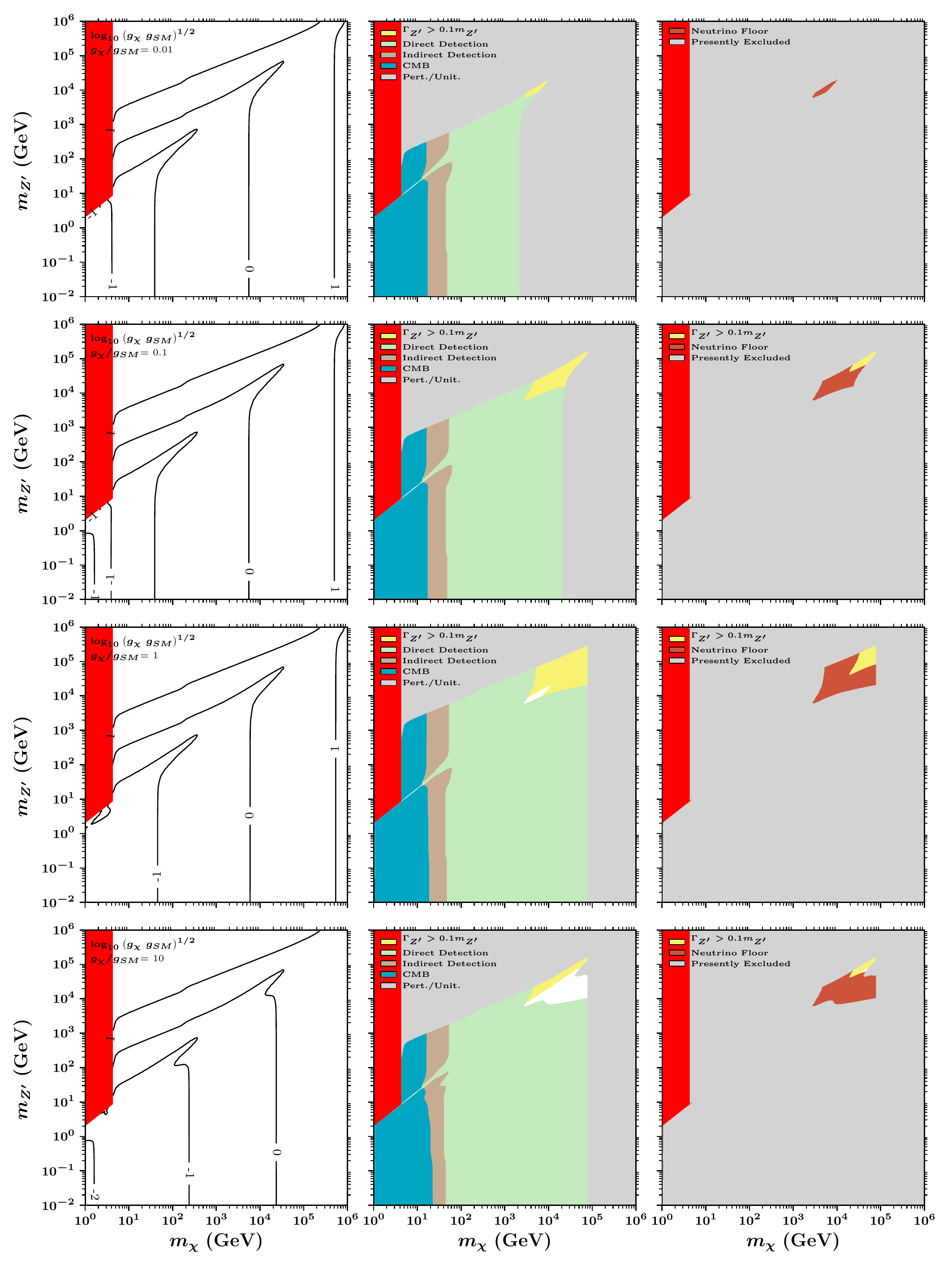}}
\caption{As in previous figures, but for the case of Dirac dark matter that is coupled to a $Z'$ with couplings only to third generation quarks. In the red regions, annihilations produce a variety of hadronic final states without velocity suppression, and are thus ruled out by a combination of CMB measurements and indirect searches.}
\label{Leptophobic_Dirac_3rd}
\end{figure}

Thus far, we have considered the case in which the $Z'$ couples equally to all SM quarks. It is, of course, plausible that different SM quarks could possess different charges under $U(1)'$, leading to non-universal effective couplings to the $Z'$. In Figs.~\ref{Leptophobic_Dirac_1st} and~\ref{Leptophobic_Dirac_3rd}, we show results for the case of Dirac dark matter that is coupled to a $Z'$ with couplings to only first or third generation quarks, respectively. In the former case, the constraints are only slightly changed, as the elastic scattering with nuclei is facilitated largely through couplings to light quarks. If the $Z'$ only couples to third generation quarks, however, the phenomenology changes in non-negligible ways. In particular, scattering with nuclei occurs through diagrams featuring heavy quark loops, leading to somewhat smaller cross sections~\cite{Agrawal:2011ze,Kaplan:1988ku,Ji:2006vx}. Furthermore, if $m_{\chi} < m_b, m_{Z'}$, the dark matter will be unable to annihilate through tree-level processes, but instead does so through loops, producing pairs of light quarks (or mesons) and leptons. If $m_\chi < m_\pi$, annihilations proceed to light leptons through an $s$-wave amplitude, a scenario that is excluded by measurements of the CMB. Between the mass of the pion and $\sim$\,$2$ GeV, a large variety of meson annihilation channels are possible, many of which are similarly excluded. Finally, between $\sim$\,$2$ GeV and the $b$-quark mass, annihilations will generate light quarks and leptons, and are again strongly constrained. Although we do not explicitly calculate the many hadronic annihilation processes that are relevant in this region of parameter space, we are confident that it is strongly excluded by CMB measurements and indirect detection. We denote this excluded region in red in Fig.~\ref{Leptophobic_Dirac_3rd}.

\subsubsection{Couplings to Leptons}

\begin{figure}[H]
\centering
\textbf{Dirac Dark Matter, Couplings to all Leptons}\par\medskip
\centerline{\includegraphics[scale=0.68]{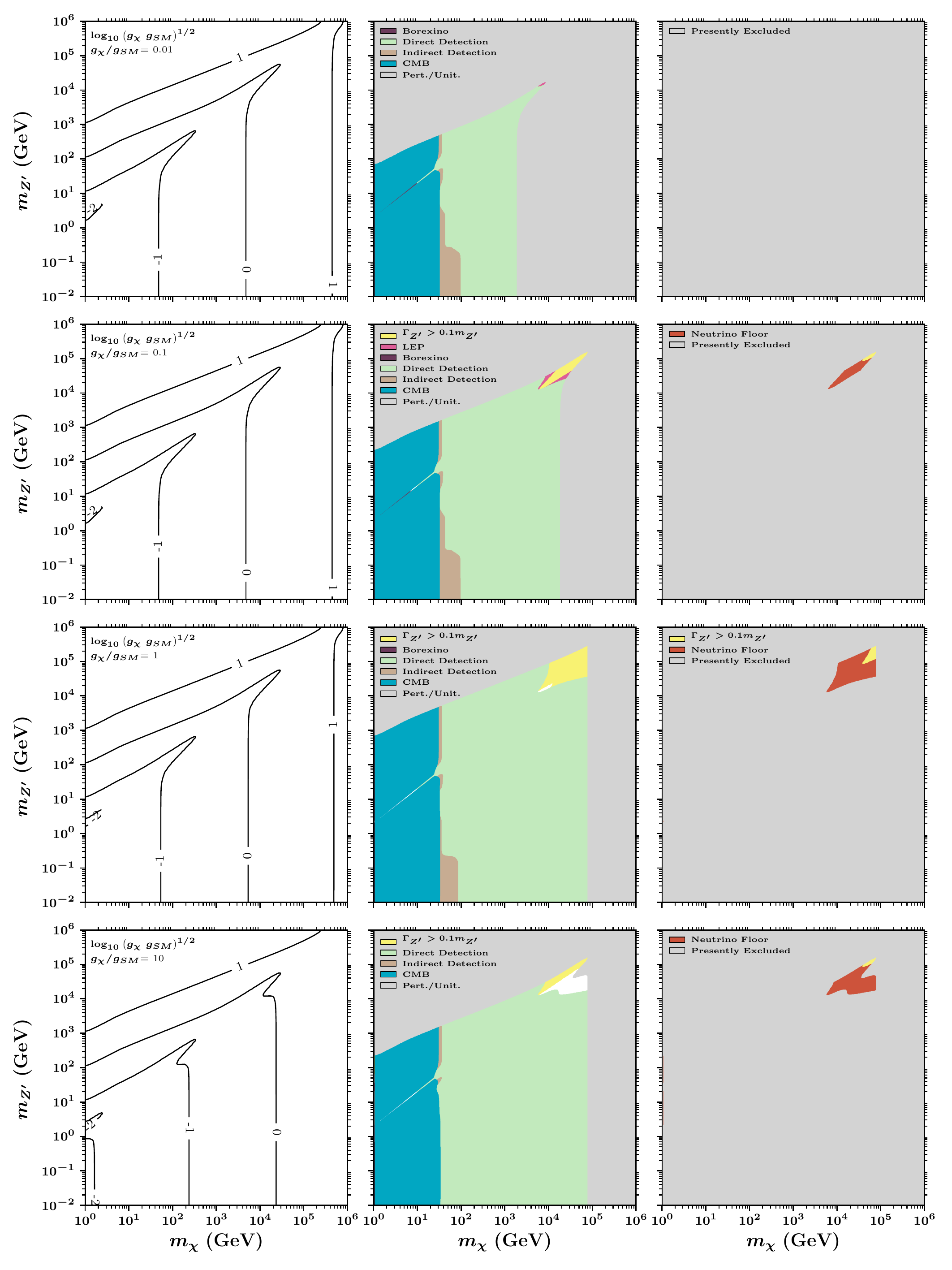}}
\caption{As in previous figures, but for the case of Dirac dark matter that is coupled to a $Z'$ with couplings to SM leptons. Again we find that the combined constraints from the cosmic microwave background, direct detection and indirect detection rule out the overwhelming majority of the parameter space shown.}
\label{Leptophilic_Dirac_All}
\end{figure}

\begin{figure}[H]
\centering
\includegraphics[width=.495\textwidth]{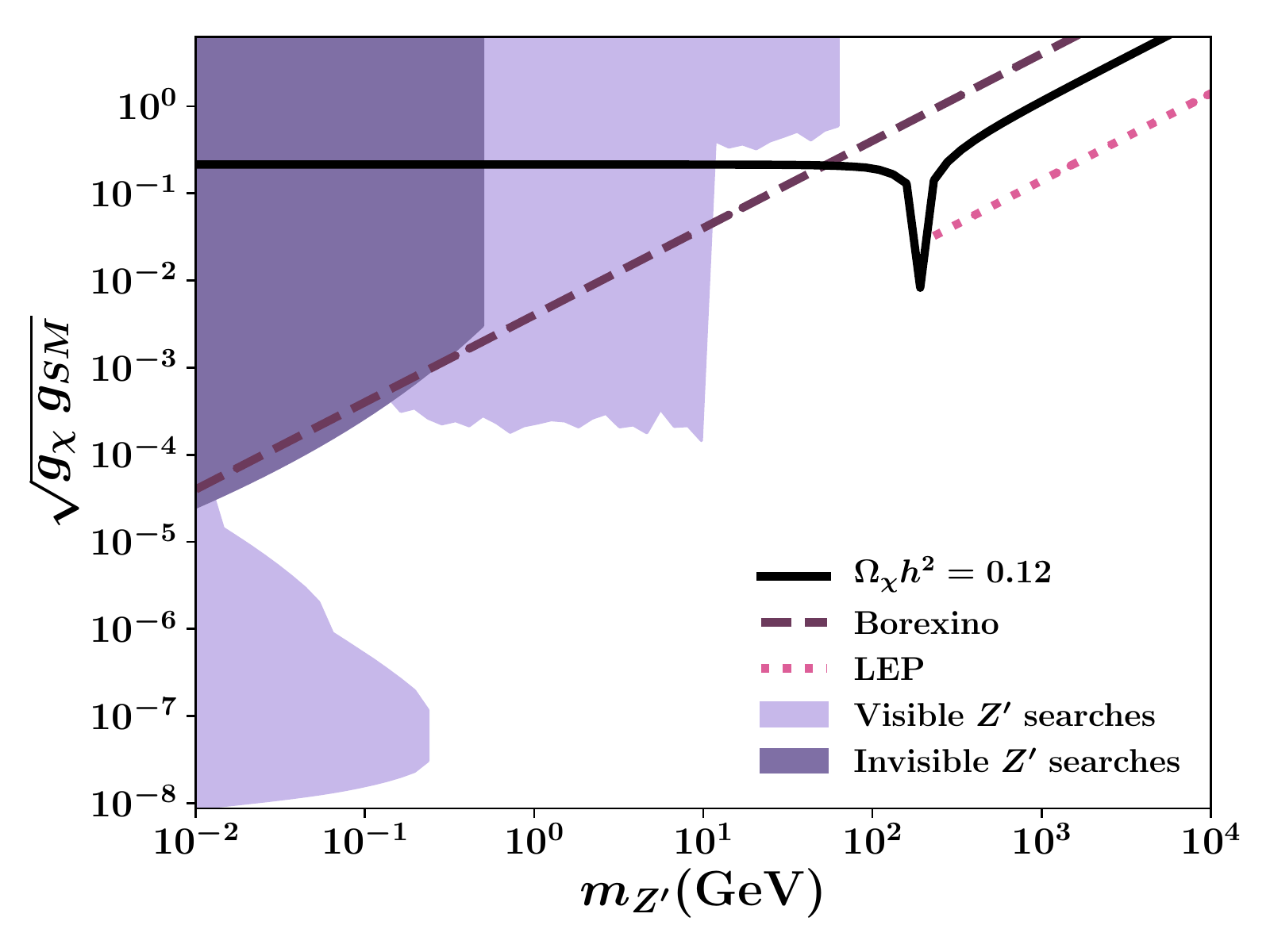}
\includegraphics[width=.495\textwidth]{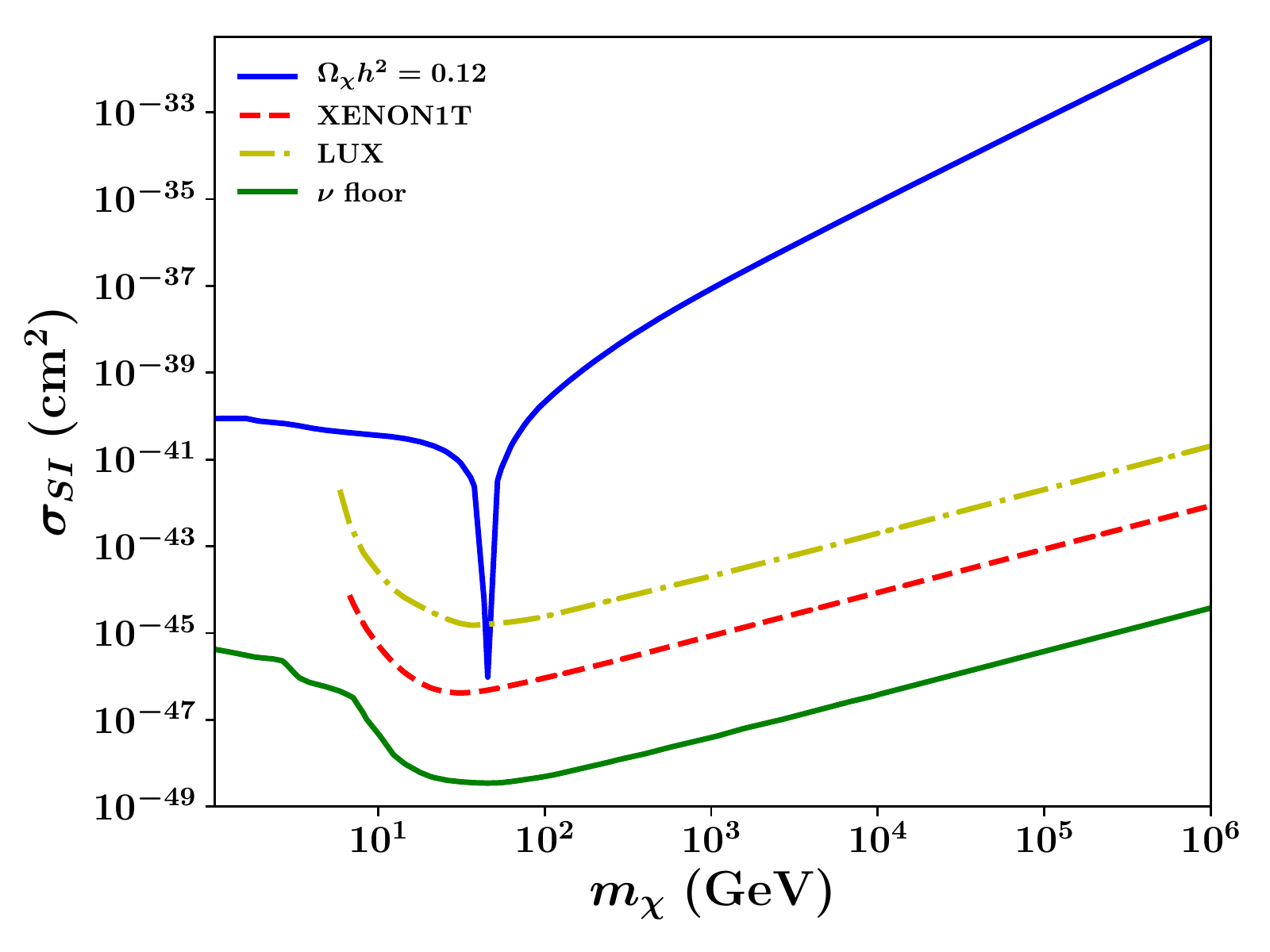}
\caption{Left frame: Constraints on Dirac dark matter that is coupled to a $Z'$ with couplings to SM leptons, for the case of $m_{\chi}=100$ GeV and $g_{\chi}/g_{\rm SM}=1$. The regions above the dotted and dashed curves are excluded by LEP and Borexino, respectively. These constraints exclude the entire range of masses in this scenario, except for a window near and slightly below resonance. Right frame:  The spin-independent elastic scattering cross section with nuclei in the same model, for the case of $m_{Z'}=100$ GeV and $g_{\chi}/g_{\rm SM}=1$. Current direct detection experiments exclude the range of models shown for all values of $m_{\chi}$ above the threshold for XENON1T, LUX and PandaX-II.}
\label{slices21}
\end{figure}

Next, we turn our attention to the case of Dirac dark matter that is coupled to a $Z'$ with couplings to SM leptons. We show our results for this case in Figs.~\ref{Leptophilic_Dirac_All} and~\ref{slices21}, where once again we find that the combined constraints from the cosmic microwave background, direct detection and indirect detection rule out the overwhelming majority of the parameter space, and that the remaining parameter space is projected to fall within the reach of future direct detection experiments. Constraints from Borexino~\cite{Harnik:2012ni} also exclude much of the parameter space in this scenario. 

It is worth noting that direct detection constraints in high mass parameter space appear to only be minimally suppressed, if at all, with respect to the scenario in which the $Z'$ couplings directly to quarks, despite the interaction being loop suppressed. We note that this is consequence of the fact that the logarithm in \Eq{epsilon} in this parameter space is quite large, given that $\Lambda = m_{Z'}/\sqrt{g_\chi g_f} \gg m_\ell$. At lower masses -- where the logarigthm is $\mathcal{O}(1)$, the direct detection bounds overlap strongly with those from indirect detection and the CMB, and thus making a straightforward comparison more difficult. 

\begin{figure}[H]
\centering
\textbf{Dirac Dark Matter, Couplings to First Generation Leptons}\par\medskip
\centerline{\includegraphics[scale=0.68]{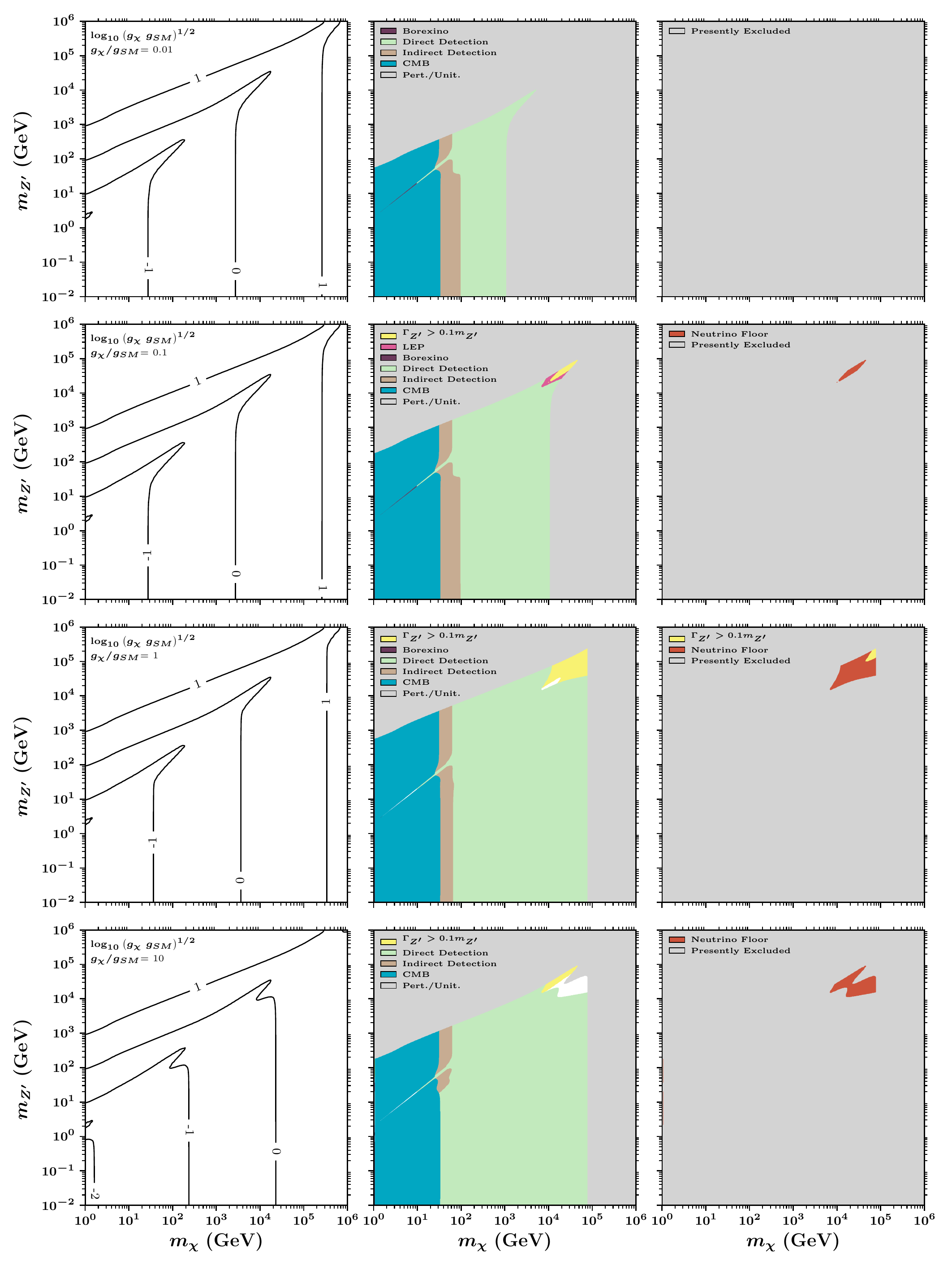}}
\caption{As in previous figures, but for the case of Dirac dark matter that is coupled to a $Z'$ with couplings only to first generation leptons. Again we find that the combined constraints from the cosmic microwave background, direct detection and indirect detection rule out the overwhelming majority of the parameter space shown.}
\label{Leptophilic_Dirac_1st}
\end{figure}
\begin{figure}[H]
\centering
\textbf{Dirac Dark Matter, Couplings to Third Generation Leptons}\par\medskip
\centerline{\includegraphics[scale=0.68]{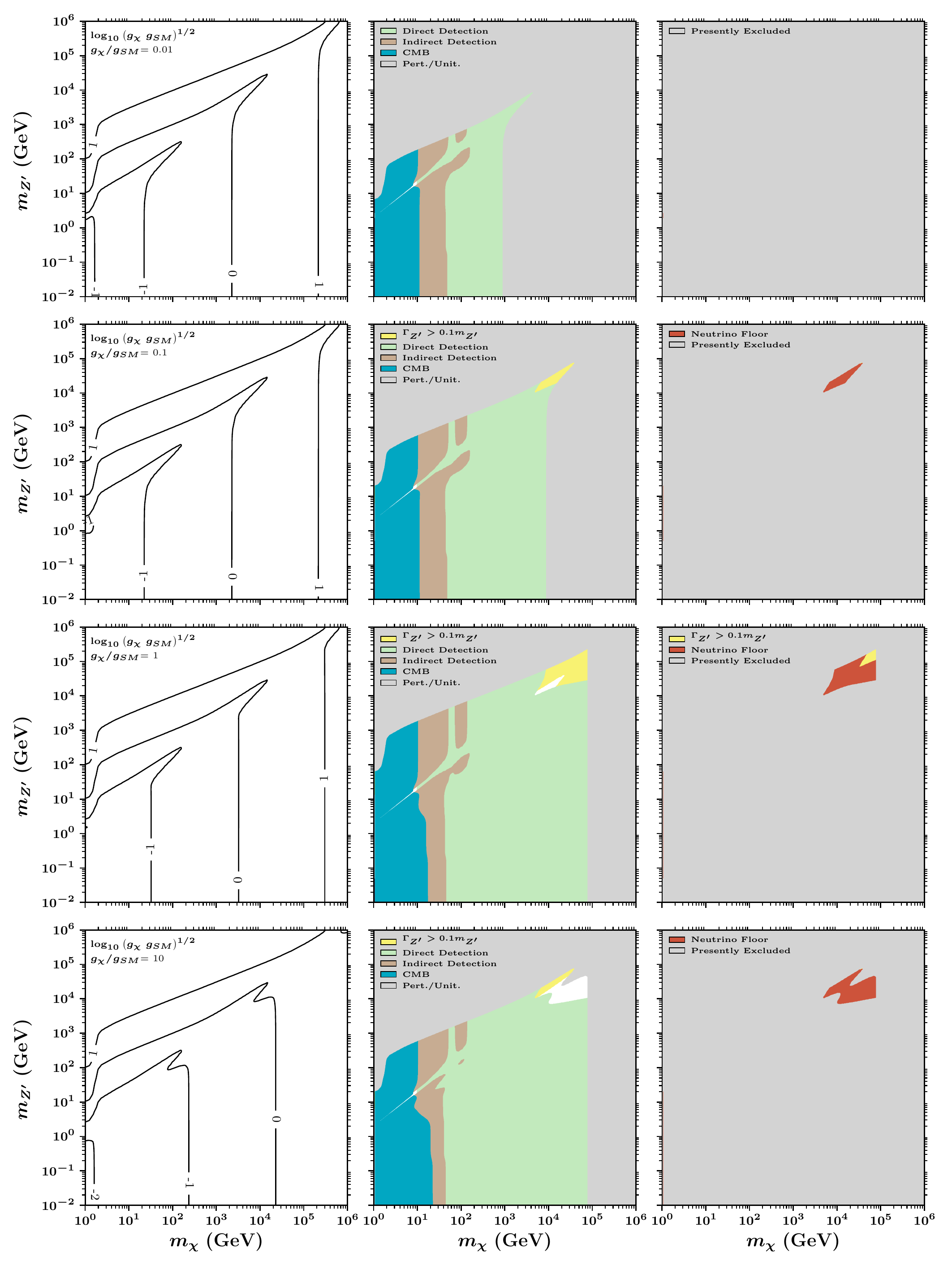}}
\caption{As in previous figures, but for the case of Dirac dark matter that is coupled to a $Z'$ with couplings only to third generation leptons. Again we find that the combined constraints from the cosmic microwave background, direct detection and indirect detection rule out the overwhelming majority of the parameter space shown.}
\label{Leptophilic_Dirac_3rd}
\end{figure}

In Figs.~\ref{Leptophilic_Dirac_1st} and~\ref{Leptophilic_Dirac_3rd}, we show results for the case of Dirac dark matter that is coupled to a $Z'$ with couplings to only first or third generation leptons, respectively. In each case, we find that the vast majority of the parameter space is currently excluded, and that future direct detection experiments are projected to cover the remaining models.

\subsection{Majorana Dark Matter}
\label{majoranaresults}

In the previous subsection, we showed that the constraints on Dirac, $Z'$ mediated dark matter leave this class of models strongly constrained across a wide range of the parameter space. In fact, such scenarios are already all but ruled out, and will be fully explored by future direct detection experiments. These constraints are much less restrictive, however, in the case of Majorana dark matter. This is true for two main reasons. First, Majorana dark matter annihilates through $p$-wave amplitudes, and is thus suppressed at low velocities, reducing the sensitivity of CMB measurements and indirect searches. For this reason, we present our results in this section for dark matter masses down to 10 MeV, below which the measurements of the primordial light element abundances exclude the parameter space. Second, the elastic scattering cross section of Majorana dark matter with nuclei is suppressed by two powers of velocity or momentum, reducing the sensitivity of direct detection experiments by a factor of approximately $\sim$\,$10^{-6}$\footnote{We note that because of this suppression, loop effects could in principle become important. This effect seems only to be relevant however for a narrow range of low mass WIMPs~\cite{Chao:2019lhb}, and thus we neglect this effect in the computations that follow.}. 

\subsubsection{Couplings to Quarks}

\begin{figure}[H]
\centering
\textbf{Majorana Dark Matter, Couplings to all Quarks}\par\medskip
\centerline{\includegraphics[scale=0.68]{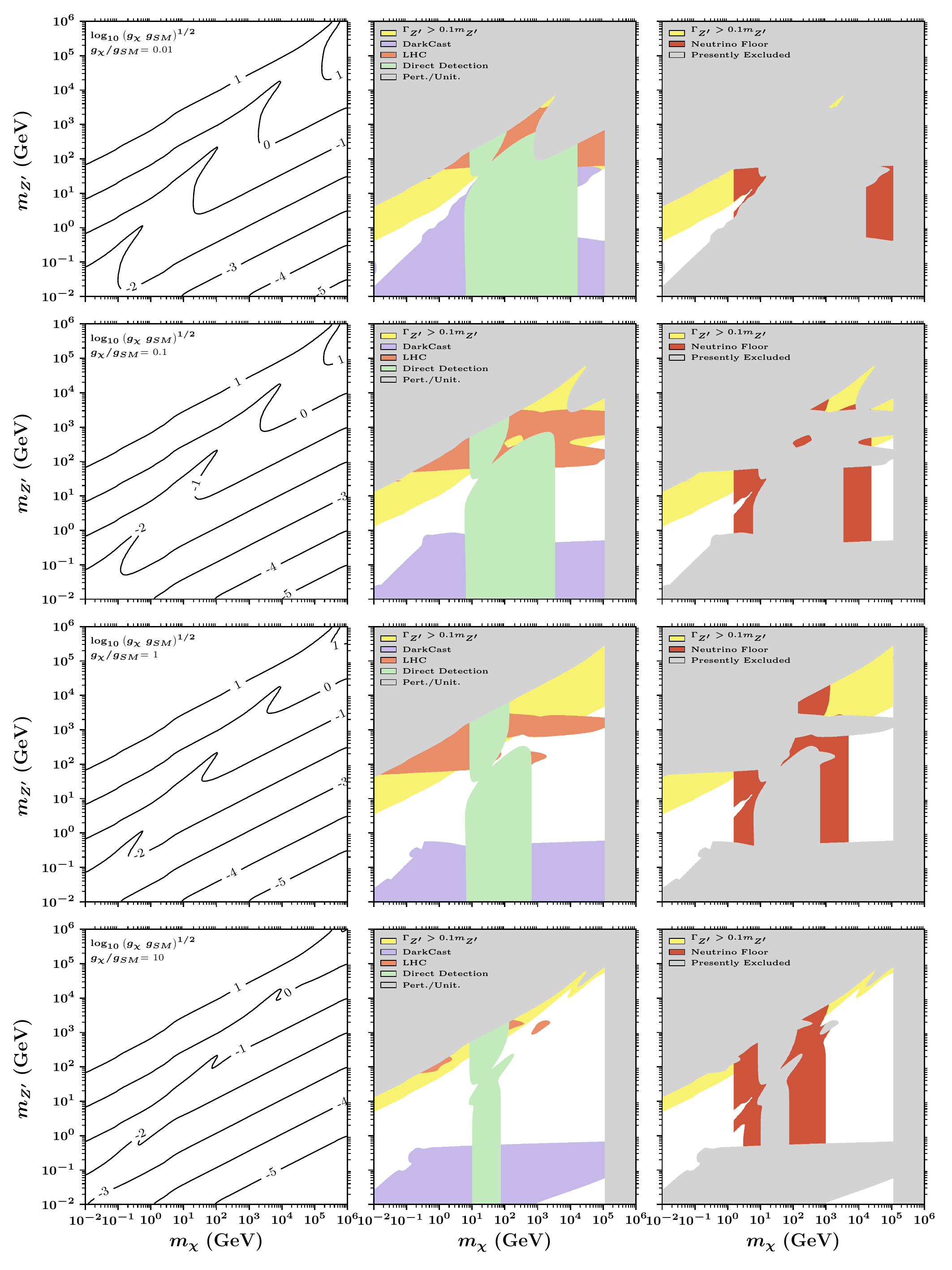}}
\caption{As in previous figures, but for the case of Majorana dark matter that is coupled to a $Z'$ with couplings to all SM quarks. Although the combined constraints from direct detection, colliders and fixed target experiments exclude significant portions of the parameter space, substantial regions remain viable. Future direct detection experiments are projected to be sensitive to much of the remaining parameter space.}
\label{Leptophobic_Majorana_All}
\end{figure}

In Fig.~\ref{Leptophobic_Majorana_All}, we show our results for the case of Majorana dark matter with a $Z'$ that couples equally to all SM quarks. In this case, constraints from direct detection, colliders and fixed target experiments exclude significant portions of the parameter space, although substantial regions remain viable (in particular for the case of $g_{\chi} \gtrsim g_{\rm SM}$). In the right frames of this figure, we see that future direct detection experiments are projected to probe a significant fraction of the remaining parameter space in this model. Even with an array of experiments that reach the neutrino floor, however, some of this parameter space will remain unexplored. 

\begin{figure}[h]
\centering
\includegraphics[width=.495\textwidth]{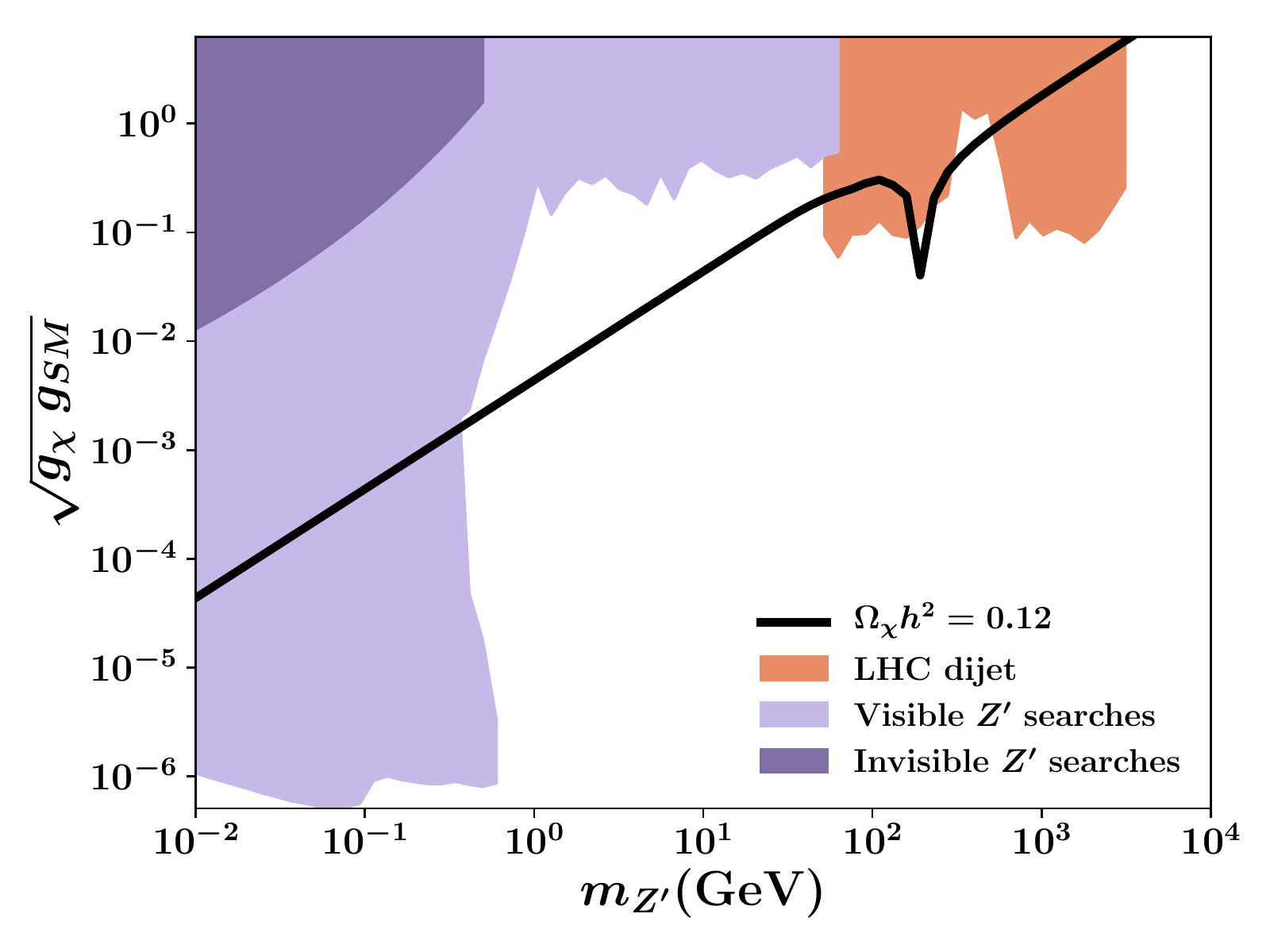}
\includegraphics[width=.495\textwidth]{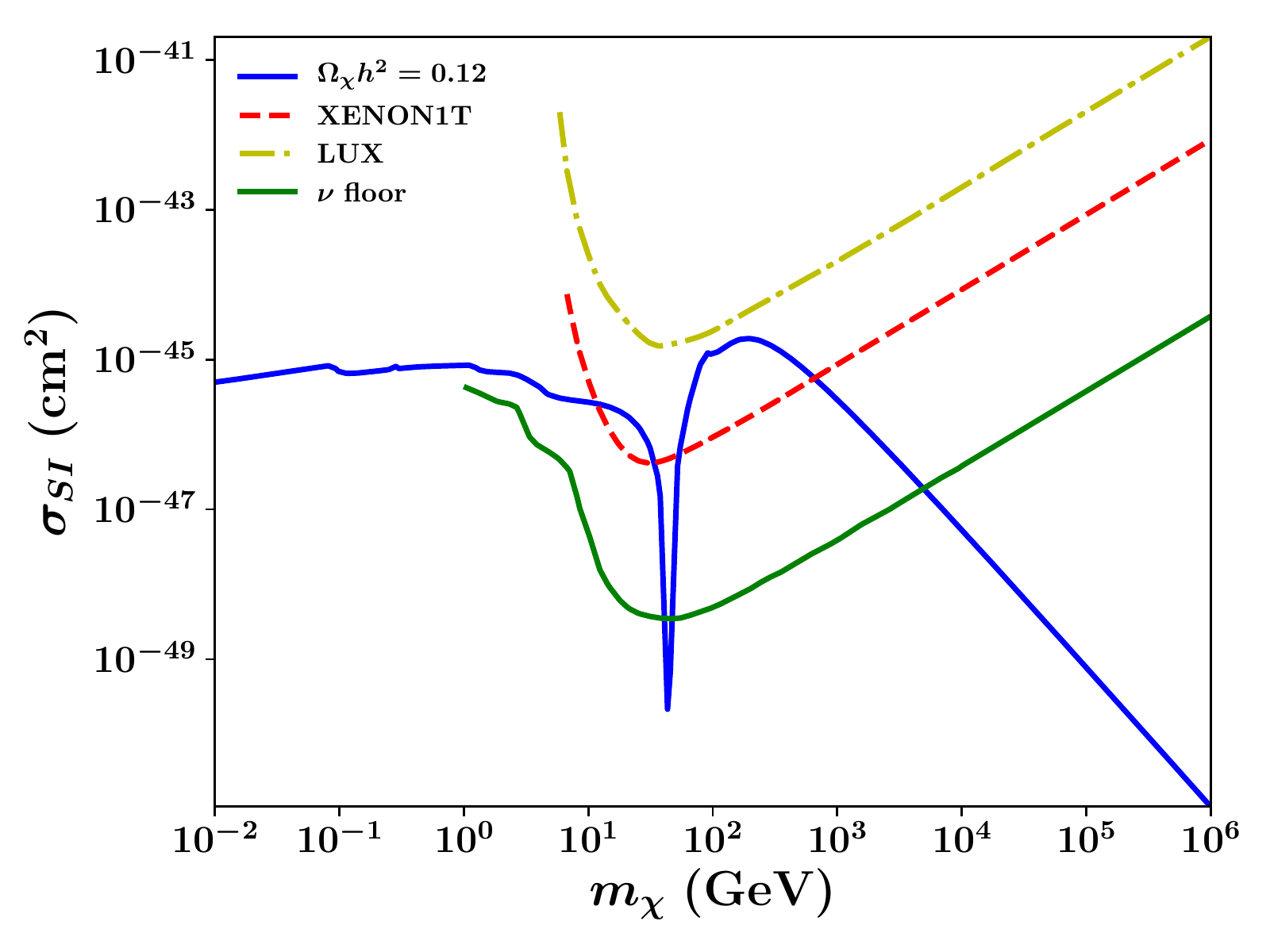}
\caption{Left Frame: Constraints on Majorana dark matter that is coupled to a $Z'$ with couplings to all SM quarks, for the case of $m_{\chi}=100$ GeV and $g_{\chi}/g_{\rm SM}=1$. Searches for light $Z'$ bosons~\cite{Ilten:2018crw} exclude $Z'$ masses below $\sim$0.4 GeV, while dijet searches at ATLAS and CMS exclude some regions of parameter space with larger values of $m_{Z'}$. Right frame: The spin-independent elastic scattering cross section with nuclei in the same model, for the case of $m_{Z'}=100$ GeV and $g_{\chi}/g_{\rm SM}=1$. Current direct detection experiments exclude dark matter with masses between $\sim$\,$10-30$ GeV and $\sim$\,$50-600$ GeV, and future direct detection experiment will explore a significantly wider range of masses.}
\label{slices31}
\end{figure}

In Fig.~\ref{slices31}, we further explore this class of models across specific slices of parameter space. For the case of $m_{\chi}=100$ GeV and $g_{\chi}/g_{\rm SM}=1$, searches for light $Z'$ bosons (as characterized using DarkCast~\cite{Ilten:2018crw}) exclude $Z'$ masses below $\sim$\,0.4 GeV, while dijet searches at ATLAS and CMS exclude some regions of parameter space with larger values of $m_{Z'}$. In the right frame, we see that for $m_{Z'}=100$ GeV and $g_{\chi}/g_{\rm SM}=1$, direct detection experiments current exclude dark matter with masses between $\sim$\,$12-33$ GeV and $\sim$\,$54-640$ GeV (although future direct detection experiment will explore a much wider range of masses).

\begin{figure}[H]
\centering
\textbf{Majorana Dark Matter, Couplings to First Generation Quarks}\par\medskip
\centerline{\includegraphics[scale=0.68]{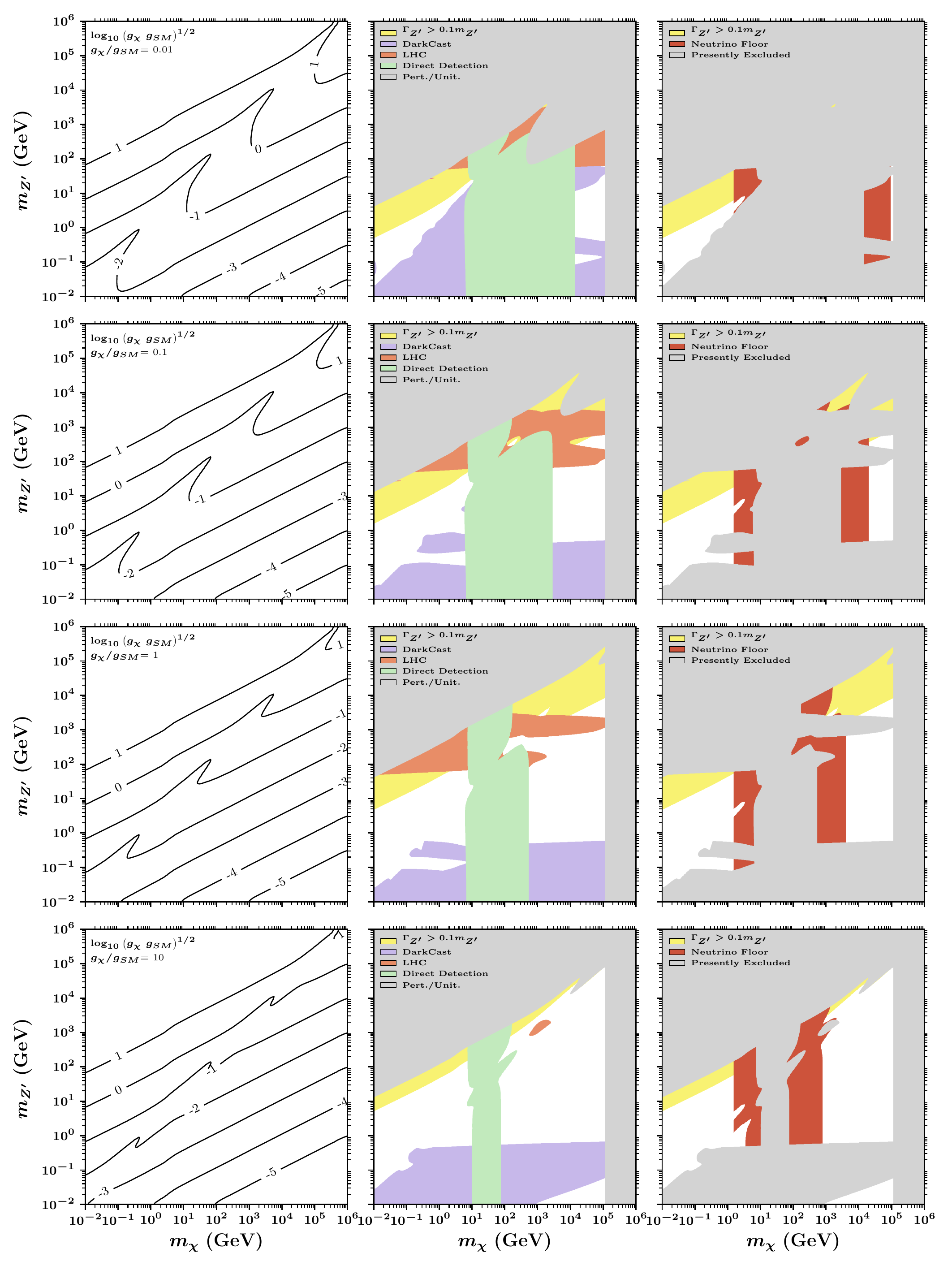}}
\caption{As in previous figures, but for the case of Majorana dark matter that is coupled to a $Z'$ with couplings only to first generation quarks. Although the combined constraints from direct detection, colliders and fixed target experiments exclude significant portions of the parameter space, substantial regions remain viable. Future direct detection experiments are projected to be sensitive to much of the remaining parameter space.}
\label{Leptophobic_Majorana_1st}
\end{figure}

\begin{figure}[H]
\centering
\textbf{Majorana Dark Matter, Couplings to Third Generation Quarks}\par\medskip
\centerline{\includegraphics[scale=0.68]{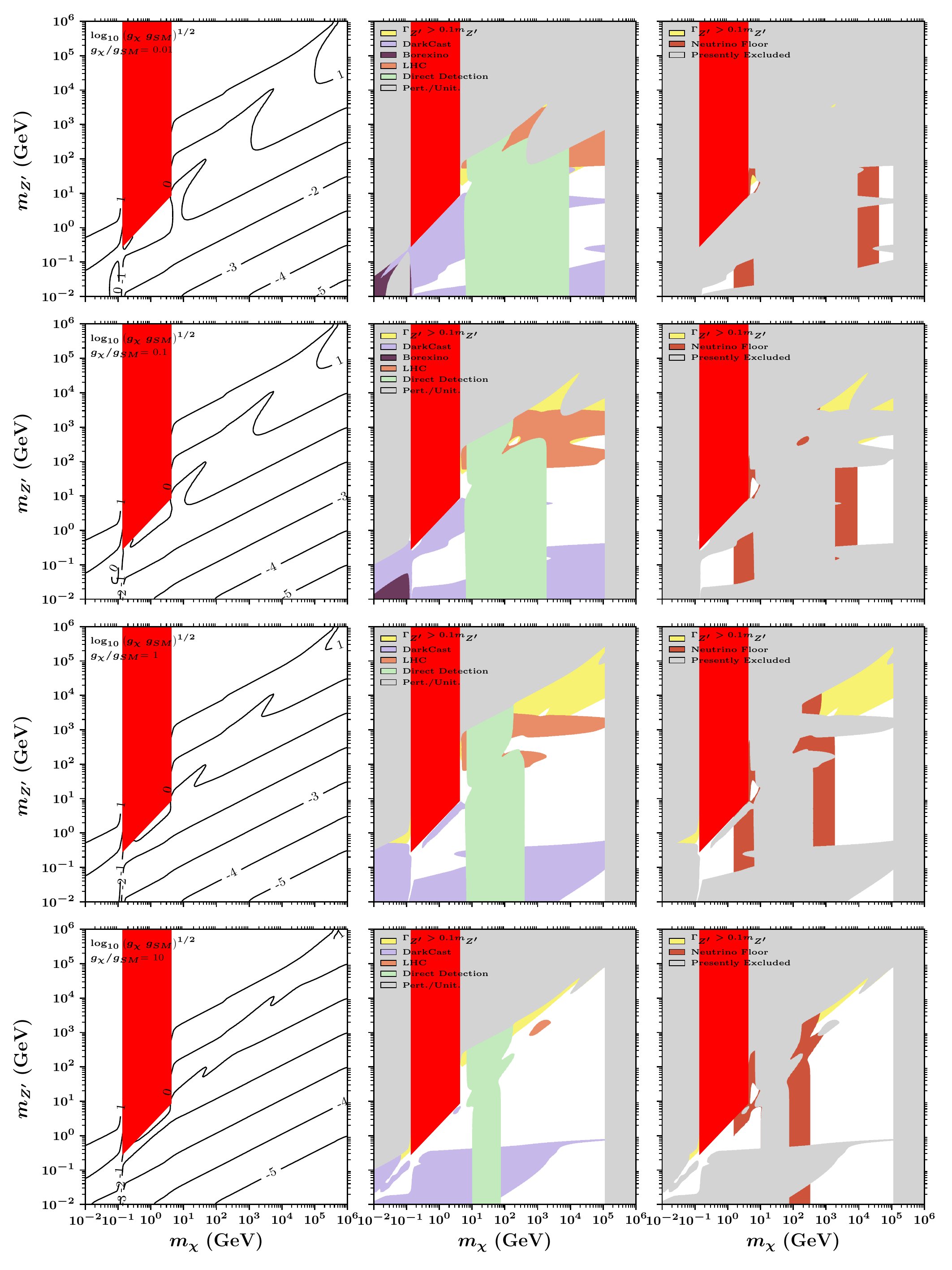}}
\caption{As in previous figures, but for the case of Majorana dark matter that is coupled to a $Z'$ with couplings only to third generation quarks. Although the combined constraints from direct detection, colliders and fixed target experiments exclude significant portions of the parameter space, substantial regions remain viable. Future direct detection experiments are projected to be sensitive to much of the remaining parameter space. In the red regions, annihilations produce a variety of hadronic final states.}
\label{Leptophobic_Majorana_3rd}
\end{figure}

In Figs.~\ref{Leptophobic_Majorana_1st} and~\ref{Leptophobic_Majorana_3rd}, we show the results for the case of Majorana dark matter that is coupled to a $Z'$ with couplings only to first or third generation quarks, respectively. Large portions of the parameter space have been (and will be) tested for models in which $g_{\rm SM} \gtrsim g_\chi$. On the other hand, when the SM coupling is suppressed, large portions of parameter space will likely remain unexplored for some time. In the third generation case, we have again blocked out in red the region of parameter space corresponding to $m_\pi < m_\chi < m_b, m_{Z^\prime}$ where dark matter annihilation occurs through loops to light quarks (or mesons) and leptons.

\subsubsection{Couplings to Leptons}

\begin{figure}[H]
\centering
\textbf{Majorana Dark Matter, Couplings to all Leptons}\par\medskip
\centerline{\includegraphics[scale=0.68]{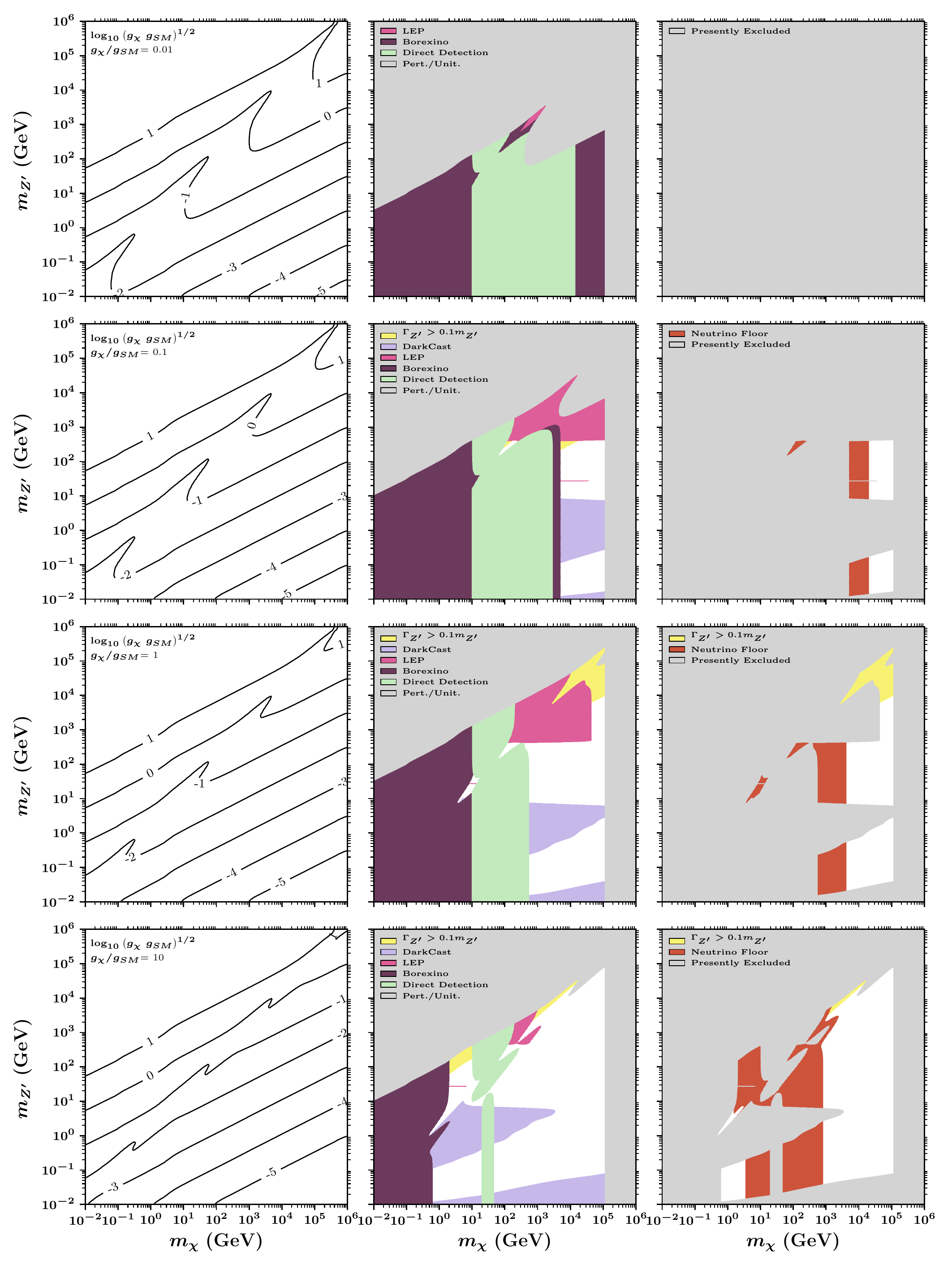}}
\caption{As in previous figures, but for the case of Majorana dark matter that is coupled to a $Z'$ with couplings to all SM leptons. Among this class of models, $Z'$ searches at Borexino, LEP, and lower energy collider and fixed target experiments provide the most powerful constraints.}
\label{Leptophilic_Majorana_All}
\end{figure}

\begin{figure}[h]
\centering
\includegraphics[width=.495\textwidth]{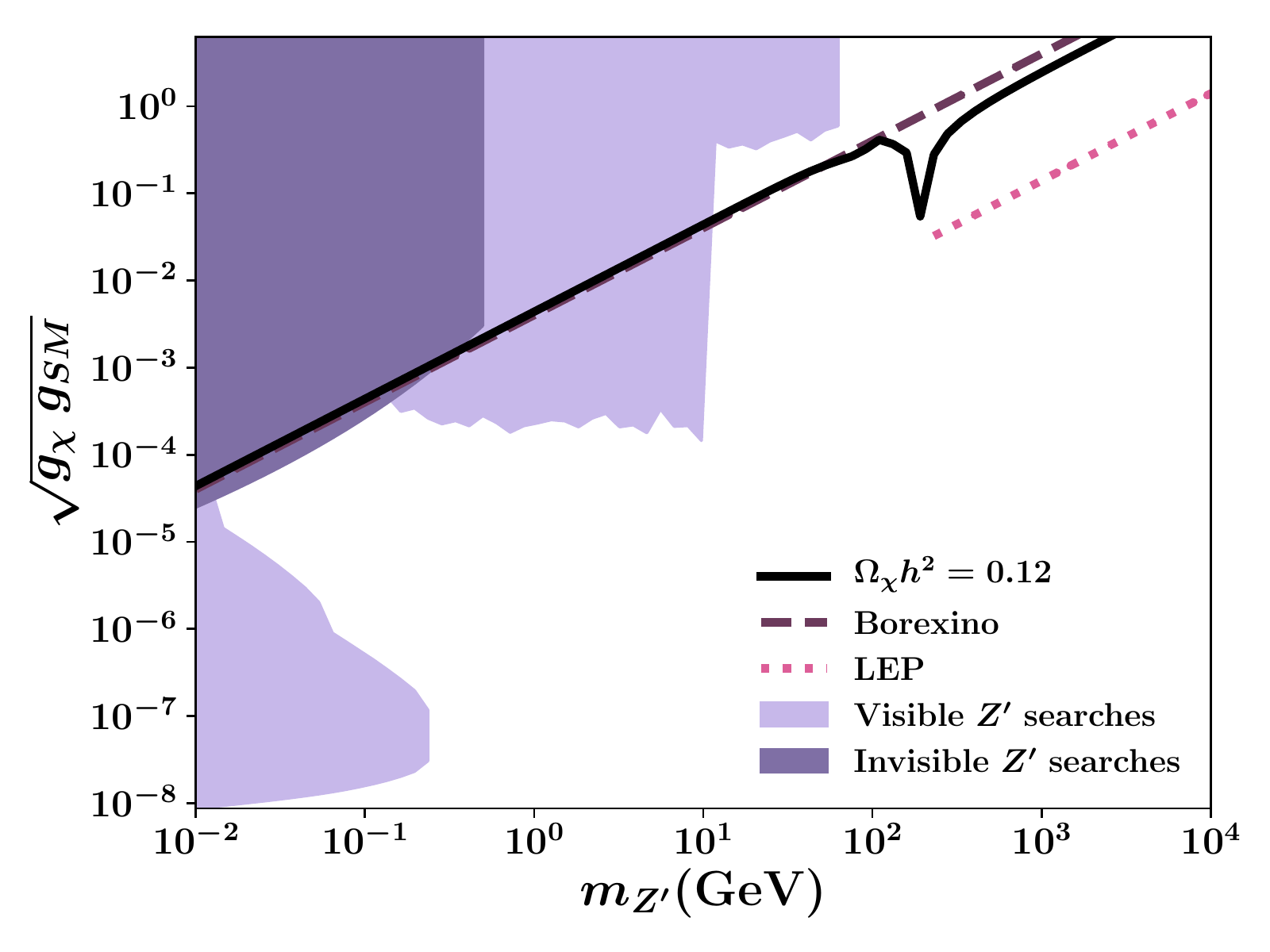}
\includegraphics[width=.495\textwidth]{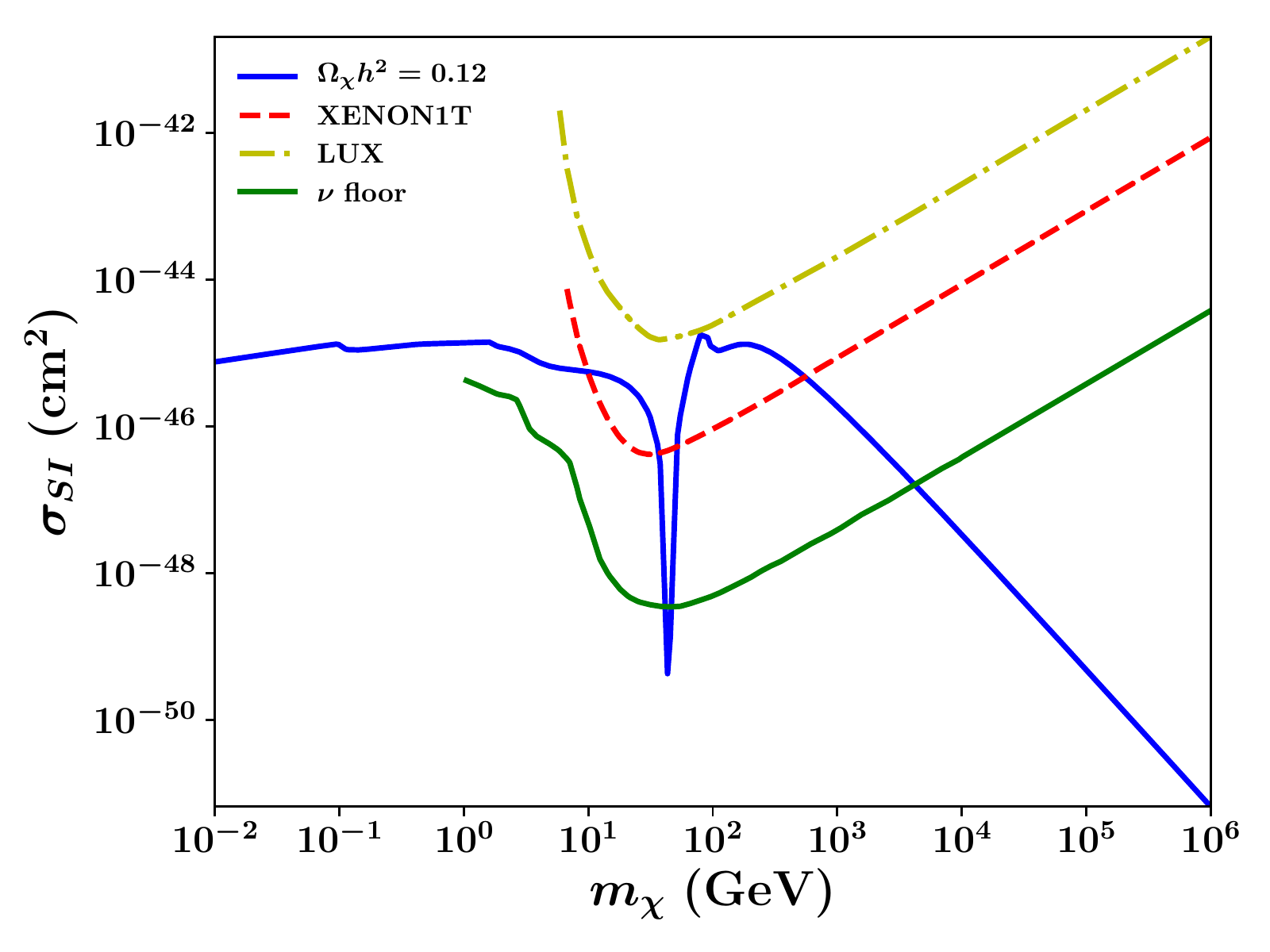}
\caption{Left Frame: Constraints on Majorana dark matter that is coupled to a $Z'$ with couplings to all SM leptons, for the case of $m_{\chi}=100$ GeV and $g_{\chi}/g_{\rm SM}=1$. The regions above the dotted and dashed curves are excluded by LEP and Borexino, respectively. Searches for light $Z'$ bosons~\cite{Ilten:2018crw} and constraints from LEP and Borexino exclude most of the parameter space shown. Right frame: The spin-independent elastic scattering cross section with nuclei in the same model, for the case of $m_{Z'}=100$ GeV and $g_{\chi}/g_{\rm SM}=1$. Current direct detection experiments exclude dark matter with masses between $\sim$\,$10-40$ GeV and $\sim$\,$50-500$ GeV, and future direct detection experiment will explore a significantly wider range of masses.}
\label{slices11}
\end{figure}

\begin{figure}[H]
\centering
\textbf{Majorana Dark Matter, Couplings to First Generation Leptons}\par\medskip
\centerline{\includegraphics[scale=0.68]{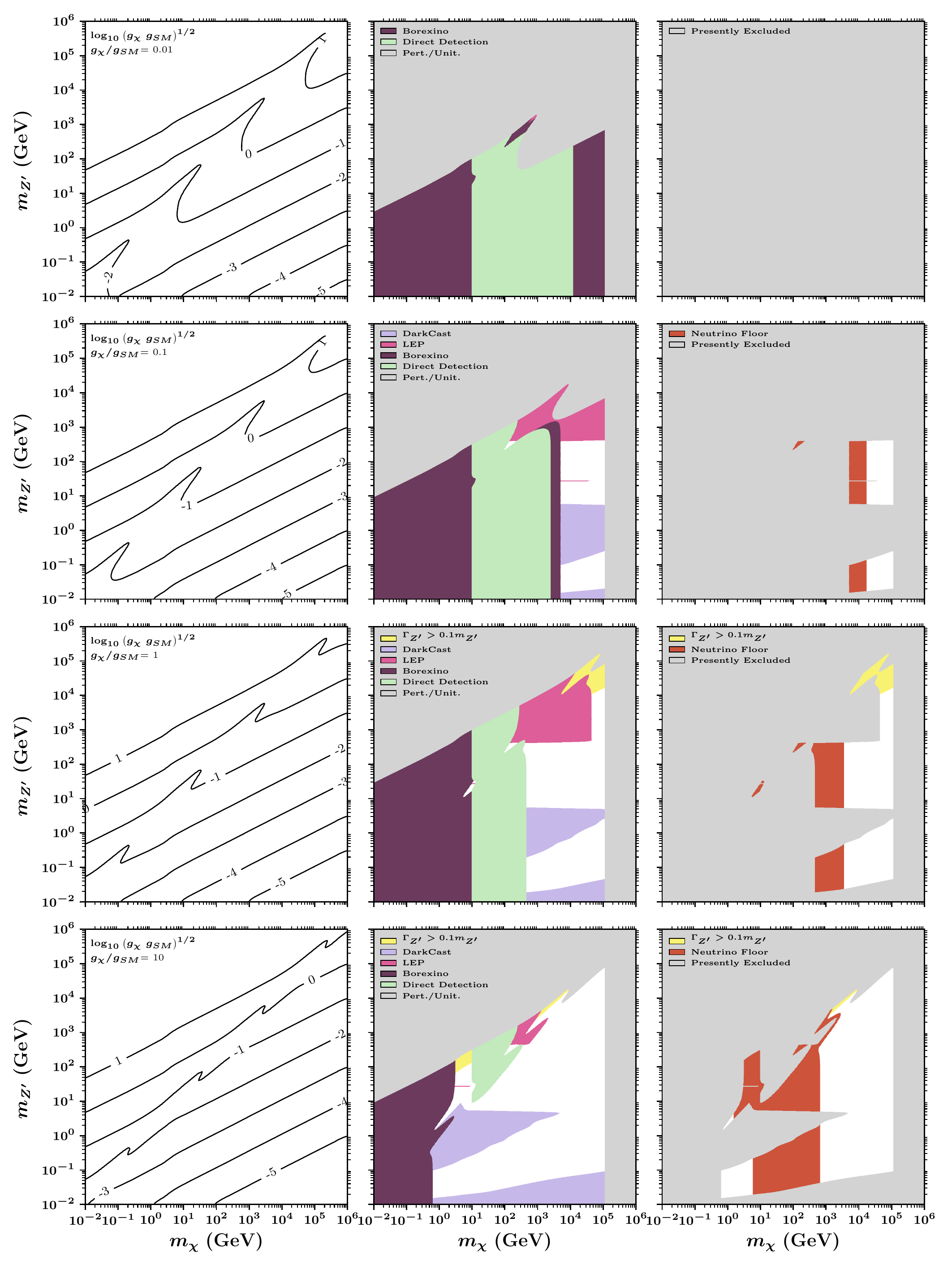}}
\caption{As in previous figures, but for the case of Majorana dark matter that is coupled to a $Z'$ with couplings only to first generation leptons. Among this class of models, $Z'$ searches at Borexino, LEP, and lower energy collider and fixed target experiments provide the most powerful constraints.}
\label{Leptophilic_Majorana_1st}
\end{figure}

\begin{figure}[H]
\centering
\textbf{Majorana Dark Matter, Couplings to Third Generation Leptons}\par\medskip
\centerline{\includegraphics[scale=0.68]{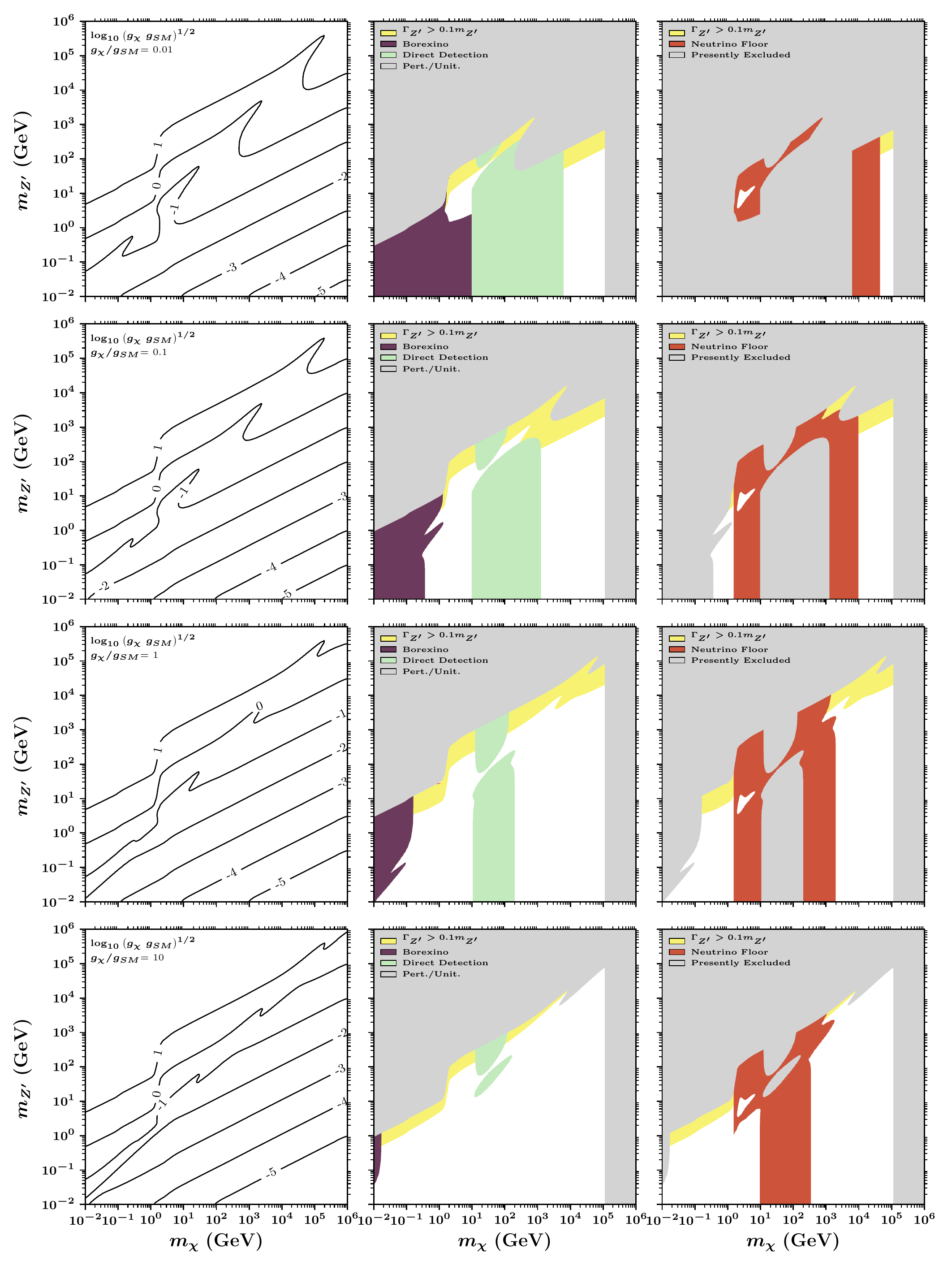}}
\caption{As in previous figures, but for the case of Majorana dark matter that is coupled to a $Z'$ with couplings only to third generation leptons. This is the least constrained class of models among those considered in this study.}
\label{Leptophilic_Majorana_3rd}
\end{figure}

Lastly, we consider the case of Majorana dark matter with a $Z'$ that couples to SM leptons. Among this class of models, $Z'$ searches at LEP, Borexino, and at lower energy colliders and fixed target experiments provide the most powerful constraints. In scenarios with couplings to all SM leptons (see Figs.~\ref{Leptophilic_Majorana_All} and~\ref{slices11}) or couplings to first generation leptons (Fig.~\ref{Leptophilic_Majorana_1st}), these constraints exclude much of the parameter space, except that with $g_{\chi} \gtrsim g_{\rm SM}$, for which this model remains largely unconstrained at high to intermediate dark matter masses. Interestingly, in such models, the loop suppressed direct detection interactions can dominate over the tree level contribution, given that the latter is $q^2$ and $v^2$ suppressed while the former is not. Future direct detection experiments will have only a modest impact on the parameter space of this model. In a scenario in which the $Z'$ couples only to third generation leptons, the constraints from $Z'$ become substantially less restrictive, as shown in Fig.~\ref{Leptophilic_Majorana_3rd}. This is because such a $Z'$ can be produced in $e^+ e^-$ collisions only through loops, and final states including electrons and muons are often easier to identify and reconstruct than those featuring tau leptons.

\section{Caveats and Theoretical Considerations}
\label{caveats}

\subsection{Models With an Axial $Z'$}

Throughout this study, we have restricted our attention to models in which the $Z'$ possesses only vectorial couplings to SM fermions. In a more generalized approach, one might also consider the possibility of non-zero axial couplings. Such generalizations generally impact the prospects for direct and indirect detection most strongly, with relatively little impact on the constraints from colliders or fixed target experiments, or on the determination of the thermal relic abundance. Regarding direct detection, adopting a purely axial coupling will lead to a spin-dependent interaction, which is suppressed relative to the spin-independent case a factor of $\sim$\,$A^2$, but that is still much larger than those found in typical Majorana models (as considered in Sec.~\ref{majoranaresults}). From the perspective of indirect detection, the low-velocity annihilation cross section is chirality suppressed in this case (\ie modified by a factor of $m_f^2 / m_\chi^2$), yielding similar indirect detection constraints as those found in Sec.~\ref{majoranaresults}. The phenomenology of an axially coupled dark matter candidate can thus be intuited from the constraints provided in this study. 

Theoretically, however, a $Z'$ with purely axial couplings to SM fermions is somewhat difficult to motivate. Although a $Z'$ will couple in a purely axial way to the dark matter if it is a Majorana fermion, the same cannot be said of SM fermions. In Ref.~\cite{Ismail:2016tod} (see also~\cite{Casas:2019edt}), the authors attempted to motivate purely axial anomaly-free models, but found that such models typically require a large number of new particles with large charge assignments. In order for these new particles to avoid experimental constraints, it is most natural to shift the characteristic mass scale of the new sector to be at or above the TeV scale, typically requiring $\mathcal{O}(1)$ gauge couplings in order to produce the correct relic abundance. The presence of a large gauge coupling and large charges implies that the running will be strong, leading Landau poles to appear at lower scales. Rather than including a new ad-hoc $U(1)$ symmetry (as done in Ref.~\cite{Ismail:2016tod}), one could instead attempt to generate purely axial couplings using a more involved symmetry breaking pattern. For example, within the context of $SO(10)$~\cite{Langacker:1980js,Langacker:2008yv,Hewett:1988xc} it is possible to arrange for a $U(1)_R \times U(1)_{B-L}$ symmetry that is unbroken at relatively low energy scales ($\Lambda \sim \text{TeV}$), and that subsequently breaks to $U(1)_Y$. The gauge boson associated with this symmetry could potentially have purely axial couplings with some SM fermions provided that the gauge couplings of $g_{B-L}$ and $g_R$ are equal (which is not generically expected to be the case). However, since the $U(1)_R$ charges of the SM up and down quarks are flipped, it is not possible in this scenario to generate purely axial interactions to either protons or neutrons, and therefore direct detections constraints will still be very restrictive (see also, Refs.~\cite{Arcadi:2017atc,Arcadi:2017hfi,King:2017cwv,Ferrari:2018rey,Camargo:2018rfi}).

\subsection{UV Complete Models}

The models considered in this study are intended to provide an adequate description of the relevant phenomenology, and do not necessarily represent a UV complete description of the underlying theory. More specifically, the gauge invariance of such a theory requires the cancellation of all anomalies arising from triangle diagrams with gauge bosons as external lines. In most cases, this requires the introduction of new chiral fermions, known as exotics~\cite{Batra:2005rh,Appelquist:2002mw}. The requirement of perturbativity implies that these particles must be lighter than approximately $m_{f} \lesssim 5.4 \, {\rm TeV} \times (m_{Z'}/100 \, {\rm GeV}) (0.1/g_{Z'}) (1/q_{\varphi})$, where $q_{\varphi}$ is the charge of the Higgs field associated with the breaking of the $U(1)'$. Such particles are constrained by the LHC and LEP~\cite{Dobrescu:2014fca}, in particular in the case of small values of $m_{Z'}$ or a large coupling. The triangle diagrams involving exotic fermions can also induce scattering processes that scale like $(E / m_{Z'})^2$, leading to stringent constraints on scenarios with a light $Z'$~\cite{Dror:2017nsg,Dror:2018wfl}.

Additionally, the $Z_2$ symmetry we have imposed by hand in order to stabilize the dark matter particle is rather arbitrary from a theoretical perspective. Many well-motivated models, however, have discrete symmetries that are a result of the symmetry breaking structure of the new gauge symmetry, or by the particle content that is needed in order to cancel anomalies (see, for example, Refs.~\cite{FileviezPerez:2010gw,Duerr:2013dza,Duerr:2014wra,Ismail:2016tod,Casas:2019edt,Ellis:2017tkh,CentellesChulia:2019gic}). Scenarios in which $Z_2$ symmetries arise naturally often involve dark matter particles with masses above the TeV scale.

\section{Implications for the WIMP Paradigm}
\label{bayesian}

In this section, we will attempt to summarize and synthesize the results of this study, in an effort to more broadly evaluate the status of the WIMP paradigm. At some level, we acknowledge that this goal is perhaps overly ambitious. The WIMP paradigm includes a vast range of models, and the collection of $Z'$ mediated scenarios considered here only begins to scratch the surface of these possibilities. That being said, this collection of models provides us with a fairly representative sample of WIMP models, and we contend that the results of this study can help to illuminate the status of the WIMP paradigm in the presence of the current constraints and projected sensitivity of direct detection, indirect detection and accelerator experiments. 

Broadly speaking, our goal here will be to quantitatively determine the extent to which experiments have probed $Z'$ mediated WIMP dark matter. This question is of course to some extent ill-defined, as someone who wholeheartedly believes that dark matter is a $\sim 10$ TeV neutralino might arrive at the conclusion that $\sim 0\%$ of the dark matter parameter space has been tested, while another convinced dark matter is particle with mass $\sim 10$ GeV having s-wave annihilations might conclude dark matter has been fully ruled out. This sentiment is a direct consequence of the individuals' prior beliefs, suggesting the question we hope to address here is best framed in the context of a Bayesian analysis. Following this intuition, for each choice of Majorana or Dirac dark matter, charge assignments, and the value of $g_{\chi}/g_{\rm SM}$, we proceed by defining a model space given by $\Theta = \{ m_{Z'}, m_{\chi}\}$, fixing the couplings using the ratio (as defined in the model) and the relic density, with each mass in $\Theta$ bounded from above by perturbativity, unitarity, and $\Gamma_{Z'} < 0.1 \,m_{Z'}$, and below by successful BBN (approximated here by $m_{\chi}, m_{Z'}> 10\,{\rm MeV}$) (see Secs.~\ref{apriori} and~\ref{cosmology}). The posterior probability given a set of observations, $X$, is then given by the following:
\begin{equation}
\label{bayes}
P(\Theta | X)= \frac{P(X | \Theta) P(\Theta)}{P(X)},
\end{equation}
where $P(X)$ is the Bayesian evidence, given by $\int \, d\Theta \, P(X|\Theta) \, P(\Theta)$, and $P(X | \Theta)$ is the likelihood. At points in parameter space that are not ruled out by the data, the value of $P(X | \Theta)$ is proportional to the volume of the parameter space that yields the measured dark matter abundance, $P(X|\theta) \propto \left(\partial \Omega_{\chi}/\partial \log_{10} (g_{\chi} \, g_{\rm SM}) \right)^{-1}$. $P(\Theta)$ is the prior on the parameters and is given by the product of priors on each parameter, $P(\Theta) = P(m_{Z'}) \times P(m_{\chi}) \times P(g_{\chi})$.

An inescapable limitation of any Bayesian analysis is the necessary reliance on intrinsically subjective priors, which can introduce biases and otherwise impact the conclusions of a study. With this in mind, we will adopt three different sets of priors, allowing the reader to weigh them as they deem appropriate. In the first case, we adopt a log-flat prior on both $m_{\chi}$ and $m_{Z'}$. While this choice may be attractive to some for its theoretical neutrality, others could be motivated by considerations such as the electroweak hierarchy problem, leading them to instead focus on scenarios that feature masses near the electroweak scale. With this in mind, our second set of priors features log-normal distributions for $m_{Z'}$ and $m_{\chi}$, centered around the mass of the SM Higgs boson ($m_h=125.1$ GeV~\cite{Aad:2015zhl}) and with a one-sigma width of one order of magnitude. Lastly, one might expect gauge couplings to generically possess values near $\mathcal{O}(0.1)$. With this in mind, we adopt in our third case priors on $(m_\chi, m_{Z'})$ such that masses requiring abnormally small couplings are deemed as `less favorable' -- specifically, after computing the relic couplings for a particular point in parameter space, the prior is given by a log-normal distribution in each coupling $g_\chi$ and $g_{\rm SM}$, centered about 0.1 and with a one-sigma width of one dex, \ie
\begin{equation}
P(m_\chi, m_{Z'}) = \frac{1}{\sqrt{2\pi} g_\chi g_{\rm SM}} e^{- \frac{(\log_{10}(g_\chi) - 0.1)^2}{2} \frac{(\log_{10}(g_{\rm SM}) - 0.1)^2}{2}}  \, ,
\end{equation}
where the functional dependence of $g_\chi$ and $g_{\rm SM}$ on $m_\chi$ and $m_{Z'}$ is understood implicitly.

\begin{table*}[t]
\centering
\begin{tabular}{|c|c||c|c|c|}
\hline
\multicolumn{2}{| c ||}{Dirac Dark Matter Model} &\multicolumn{3}{ c |}{Prior} \\
\hline
$Z'$ Couples To   & $g_{\chi}/g_{\rm SM}$     & Log-Flat & $m_{\chi}, m_{Z'} \sim \mathcal{O}(m_h)$ & $g_{\chi}, g_{\rm SM} \sim \mathcal{O}(0.1)$  \\
\hline                          
 \multicolumn{2}{| c ||}{} & $P_{\rm current}$ \, $P_{\rm future}$         & $P_{\rm current}$ \, $P_{\rm future}$   & $P_{\rm current}$ \, $P_{\rm future}$   \\
 \hline
 All Quarks &0.01 & 0.00  \,\,\,\,\,\,\,\,\,\, 0.00       & 0.00  \,\,\,\,\,\,\,\,\,\, 0.00       & 0.00  \,\,\,\,\,\,\,\,\,\, 0.00        \\
  			&0.1  & 0.00       \,\,\,\,\,\,\,\,\,\, 0.00       & 0.00  \,\,\,\,\,\,\,\,\,\, 0.00       & 0.00  \,\,\,\,\,\,\,\,\,\, 0.00        \\
  				&1.0  & 0.00       \,\,\,\,\,\,\,\,\,\, 0.00       & 0.00  \,\,\,\,\,\,\,\,\,\, 0.00         & 0.00  \,\,\,\,\,\,\,\,\,\, 0.00        \\
				 & 10.0 & 0.02    \,\,\,\,\,\,\,\,\,\, 0.00       & 0.02  \,\,\,\,\,\,\,\,\,\, 0.00       & 0.01  \,\,\,\,\,\,\,\,\,\, 0.00        \\
 \hline
 1st Gen.~Quarks &0.01 & 0.00  \,\,\,\,\,\,\,\,\,\, 0.00       & 0.00  \,\,\,\,\,\,\,\,\,\, 0.00       & 0.00  \,\,\,\,\,\,\,\,\,\, 0.00        \\
  			&0.1  & 0.00       \,\,\,\,\,\,\,\,\,\, 0.00       & 0.00  \,\,\,\,\,\,\,\,\,\, 0.00       & 0.00  \,\,\,\,\,\,\,\,\,\, 0.00        \\
  				&1.0  & 0.00       \,\,\,\,\,\,\,\,\,\, 0.00       & 0.00  \,\,\,\,\,\,\,\,\,\, 0.00         & 0.00  \,\,\,\,\,\,\,\,\,\, 0.00        \\
				 & 10.0 & 0.01    \,\,\,\,\,\,\,\,\,\, 0.00       & 0.01  \,\,\,\,\,\,\,\,\,\, 0.00       & 0.01  \,\,\,\,\,\,\,\,\,\, 0.00        \\
 \hline
 3rd Gen.~Quarks &0.01 & 0.00  \,\,\,\,\,\,\,\,\,\, 0.00       & 0.00  \,\,\,\,\,\,\,\,\,\, 0.00       & 0.00  \,\,\,\,\,\,\,\,\,\, 0.00        \\
  			&0.1  & 0.00       \,\,\,\,\,\,\,\,\,\, 0.00       & 0.00  \,\,\,\,\,\,\,\,\,\, 0.00       & 0.00  \,\,\,\,\,\,\,\,\,\, 0.00        \\
  				&1.0  & 0.00       \,\,\,\,\,\,\,\,\,\, 0.00       & 0.00  \,\,\,\,\,\,\,\,\,\, 0.00         & 0.00  \,\,\,\,\,\,\,\,\,\, 0.00        \\
				 & 10.0 & 0.01    \,\,\,\,\,\,\,\,\,\, 0.00       & 0.01  \,\,\,\,\,\,\,\,\,\, 0.00       & 0.01  \,\,\,\,\,\,\,\,\,\, 0.00        \\
 \hline
  All Leptons &0.01 & 0.00  \,\,\,\,\,\,\,\,\,\, 0.00       & 0.00  \,\,\,\,\,\,\,\,\,\, 0.00       & 0.00  \,\,\,\,\,\,\,\,\,\, 0.00        \\
  			&0.1  & 0.00       \,\,\,\,\,\,\,\,\,\, 0.00       & 0.00  \,\,\,\,\,\,\,\,\,\, 0.00       & 0.00  \,\,\,\,\,\,\,\,\,\, 0.00        \\
  				&1.0  & 0.00       \,\,\,\,\,\,\,\,\,\, 0.00       & 0.00  \,\,\,\,\,\,\,\,\,\, 0.00         & 0.00  \,\,\,\,\,\,\,\,\,\, 0.00        \\
				 & 10.0 & 0.02    \,\,\,\,\,\,\,\,\,\, 0.00       & 0.02  \,\,\,\,\,\,\,\,\,\, 0.00       & 0.02  \,\,\,\,\,\,\,\,\,\, 0.00        \\
 \hline
 1st Gen.~Leptons &0.01 & 0.00  \,\,\,\,\,\,\,\,\,\, 0.00       & 0.00  \,\,\,\,\,\,\,\,\,\, 0.00       & 0.00  \,\,\,\,\,\,\,\,\,\, 0.00        \\
  			&0.1  & 0.00       \,\,\,\,\,\,\,\,\,\, 0.00       & 0.00  \,\,\,\,\,\,\,\,\,\, 0.00       & 0.00  \,\,\,\,\,\,\,\,\,\, 0.00        \\
  				&1.0  & 0.00       \,\,\,\,\,\,\,\,\,\, 0.00       & 0.00  \,\,\,\,\,\,\,\,\,\, 0.00         & 0.00  \,\,\,\,\,\,\,\,\,\, 0.00        \\
				 & 10.0 & 0.01    \,\,\,\,\,\,\,\,\,\, 0.00       & 0.01  \,\,\,\,\,\,\,\,\,\, 0.00       & 0.02  \,\,\,\,\,\,\,\,\,\, 0.00        \\
 \hline
  3rd Gen.~Leptons &0.01 & 0.00  \,\,\,\,\,\,\,\,\,\, 0.00       & 0.00  \,\,\,\,\,\,\,\,\,\, 0.00       & 0.00  \,\,\,\,\,\,\,\,\,\, 0.00        \\
  			&0.1  & 0.00       \,\,\,\,\,\,\,\,\,\, 0.00       & 0.00  \,\,\,\,\,\,\,\,\,\, 0.00      & 0.00  \,\,\,\,\,\,\,\,\,\, 0.00        \\
  				&1.0  & 0.00       \,\,\,\,\,\,\,\,\,\, 0.00       & 0.00  \,\,\,\,\,\,\,\,\,\, 0.00         & 0.00  \,\,\,\,\,\,\,\,\,\, 0.00        \\
				 & 10.0 & 0.01    \,\,\,\,\,\,\,\,\,\, 0.00      & 0.01  \,\,\,\,\,\,\,\,\,\, 0.00       & 0.02  \,\,\,\,\,\,\,\,\,\, 0.00        \\
    \hline
\end{tabular}
\caption{The current probabilities ($P_{\rm current}$) and the projected probabilities in lieu of any detection by an array of neutrino-floor direct detection experiments ($P_{\rm future}$), for the case of Dirac, $Z'$ mediated dark matter. We present results corresponding to three sets of Bayesian priors, as described in the text. In this case, the vast majority of the parameter space is already ruled out, and the little remaining viable parameter space will be tested by upcoming direct detection experiments.}     
\label{bayestable2}
\end{table*}

\begin{table*}[t]
\centering
\begin{tabular}{|c|c||c|c|c|}
\hline
\multicolumn{2}{| c ||}{ Majorana Dark Matter Model} &\multicolumn{3}{ c |}{Prior} \\
\hline
$Z'$ Couples To   & $g_{\chi}/g_{\rm SM}$     & Log-Flat & $m_{\chi}, m_{Z'} \sim \mathcal{O}(m_h)$ & $g_{\chi}, g_{\rm SM} \sim \mathcal{O}(0.1)$  \\
\hline                          
 \multicolumn{2}{| c ||}{} & $P_{\rm current}$ \, $P_{\rm future}$         & $P_{\rm current}$ \, $P_{\rm future}$   & $P_{\rm current}$ \, $P_{\rm future}$   \\
 \hline
 All Quarks &0.01 & 0.13  \,\,\,\,\,\,\,\,\,\, 0.06       & 0.07  \,\,\,\,\,\,\,\,\,\, 0.01       & 0.11  \,\,\,\,\,\,\,\,\,\, 0.08        \\
  			&0.1  & 0.27       \,\,\,\,\,\,\,\,\,\, 0.18       & 0.23  \,\,\,\,\,\,\,\,\,\, 0.05       & 0.45  \,\,\,\,\,\,\,\,\,\, 0.29        \\
  				&1.0  & 0.37       \,\,\,\,\,\,\,\,\,\, 0.27       & 0.40  \,\,\,\,\,\,\,\,\,\, 0.10         & 0.64  \,\,\,\,\,\,\,\,\,\, 0.44        \\
				 & 10.0 & 0.43    \,\,\,\,\,\,\,\,\,\, 0.33       & 0.73  \,\,\,\,\,\,\,\,\,\, 0.27       & 0.76  \,\,\,\,\,\,\,\,\,\, 0.61        \\
 \hline
 1st Gen.~Quarks &0.01 & 0.15  \,\,\,\,\,\,\,\,\,\, 0.08       & 0.07  \,\,\,\,\,\,\,\,\,\, 0.01       & 0.13  \,\,\,\,\,\,\,\,\,\, 0.09        \\
  				&0.1  & 0.32       \,\,\,\,\,\,\,\,\,\, 0.21       & 0.22  \,\,\,\,\,\,\,\,\,\, 0.05       & 0.45  \,\,\,\,\,\,\,\,\,\, 0.30        \\
  				&1.0  & 0.40       \,\,\,\,\,\,\,\,\,\, 0.29       & 0.44  \,\,\,\,\,\,\,\,\,\, 0.12         & 0.69  \,\,\,\,\,\,\,\,\,\, 0.47        \\
				 & 10.0 & 0.43    \,\,\,\,\,\,\,\,\,\, 0.34       & 0.71  \,\,\,\,\,\,\,\,\,\, 0.29       & 0.73  \,\,\,\,\,\,\,\,\,\, 0.60        \\
 \hline
 3rd Gen.~Quarks &0.01 & 0.19  \,\,\,\,\,\,\,\,\,\, 0.10       & 0.11  \,\,\,\,\,\,\,\,\,\, 0.04       & 0.17  \,\,\,\,\,\,\,\,\,\, 0.15        \\
  				&0.1  & 0.29       \,\,\,\,\,\,\,\,\,\, 0.20       & 0.22  \,\,\,\,\,\,\,\,\,\, 0.08       & 0.48  \,\,\,\,\,\,\,\,\,\, 0.41        \\
  				&1.0  & 0.42       \,\,\,\,\,\,\,\,\,\, 0.31       & 0.54  \,\,\,\,\,\,\,\,\,\, 0.20         & 0.75  \,\,\,\,\,\,\,\,\,\, 0.55        \\
				 & 10.0 & 0.52    \,\,\,\,\,\,\,\,\,\, 0.45       & 0.70  \,\,\,\,\,\,\,\,\,\, 0.44       & 0.70  \,\,\,\,\,\,\,\,\,\, 0.59        \\
 \hline
  All Leptons &0.01 & 0.00  \,\,\,\,\,\,\,\,\,\, 0.00       & 0.00  \,\,\,\,\,\,\,\,\,\, 0.00       & 0.00  \,\,\,\,\,\,\,\,\,\, 0.00        \\
  				&0.1  & 0.16  \,\,\,\,\,\,\,\,\,\, 0.09       & 0.08  \,\,\,\,\,\,\,\,\,\, 0.02       & 0.14  \,\,\,\,\,\,\,\,\,\, 0.08        \\
				 &1.0  & 0.33       \,\,\,\,\,\,\,\,\,\, 0.21       & 0.42  \,\,\,\,\,\,\,\,\,\, 0.09         & 0.44  \,\,\,\,\,\,\,\,\,\, 0.22        \\
				 & 10.0 & 0.54    \,\,\,\,\,\,\,\,\,\, 0.34       & 0.83  \,\,\,\,\,\,\,\,\,\, 0.30       & 0.54  \,\,\,\,\,\,\,\,\,\, 0.38        \\
 \hline
 1st Gen.~Leptons &0.01 & 0.00  \,\,\,\,\,\,\,\,\,\, 0.00       & 0.00  \,\,\,\,\,\,\,\,\,\, 0.00       & 0.00  \,\,\,\,\,\,\,\,\,\, 0.00        \\
  				&0.1  & 0.17  \,\,\,\,\,\,\,\,\,\, 0.11       & 0.08  \,\,\,\,\,\,\,\,\,\, 0.03       & 0.15  \,\,\,\,\,\,\,\,\,\, 0.09        \\
				 &1.0  & 0.33       \,\,\,\,\,\,\,\,\,\, 0.22       & 0.43  \,\,\,\,\,\,\,\,\,\, 0.11         & 0.41  \,\,\,\,\,\,\,\,\,\, 0.23        \\
				 & 10.0 & 0.53    \,\,\,\,\,\,\,\,\,\, 0.33       & 0.84  \,\,\,\,\,\,\,\,\,\, 0.32       & 0.47  \,\,\,\,\,\,\,\,\,\, 0.34        \\
 \hline
  3rd Gen.~Leptons &0.01 & 0.25  \,\,\,\,\,\,\,\,\,\, 0.07       & 0.33  \,\,\,\,\,\,\,\,\,\, 0.03       & 0.24  \,\,\,\,\,\,\,\,\,\, 0.04        \\
  				&0.1  & 0.50       \,\,\,\,\,\,\,\,\,\, 0.27       & 0.36  \,\,\,\,\,\,\,\,\,\, 0.06       & 0.37  \,\,\,\,\,\,\,\,\,\, 0.17        \\
  				&1.0  & 0.69       \,\,\,\,\,\,\,\,\,\, 0.44       & 0.72  \,\,\,\,\,\,\,\,\,\, 0.18         & 0.64  \,\,\,\,\,\,\,\,\,\, 0.39        \\
				 & 10.0 & 0.81    \,\,\,\,\,\,\,\,\,\, 0.61       & 0.94  \,\,\,\,\,\,\,\,\,\, 0.42       & 0.70  \,\,\,\,\,\,\,\,\,\, 0.53        \\
  \hline
\end{tabular}
\caption{As in Table~\ref{bayestable2}, but for the case of Majorana, $Z'$ mediated dark matter. With the exception of those models with large couplings to first generation leptons, $Z'$ mediated Majorana dark matter models tend to feature current probabilities in the range of $\mathcal{O}(0.1-0.8)$, with prospects for significant improvement from upcoming direct detection experiments.}     
\label{bayestable1}
\end{table*}

As mentioned, we would like to determine the fraction of parameter space within each model which to-date has been `ruled out', defined here to mean that for a particular choice of $m_\chi$ and $m_{Z'}$ experimental observations constrain this candidate (at the CLs defined in \Sec{pheno}) from accounting for the entirety of the dark matter. To this end, for each model and choice of prior, we define the ratio of the Bayesian evidence computed using current experimental constraints to the Bayesian evidence in the absence of any constraint, \ie
\begin{equation}\label{eq:pcur}
P_{\rm current} = \frac{\int \, d\Theta \, P(X_{\rm current}|\Theta) \, P(\Theta)}{\int \, d\Theta \, P(X_{\rm pre-bounds}|\Theta) \, P(\Theta)} \, .
\end{equation}
At first glance it may not obvious why $P_{\rm current}$, which is not a well-defined Bayesian statistic, should be thought of a probability related to the fractional parameter space excluded by current experiments. To understand the significance of $P_{\rm current}$ in terms of well-defined Bayesian quantities, we will consider Bayes' factor $B$, defined by the ratio of the Bayesian evidence in two competing models. Specifically, we will consider comparing the Bayes' factor between one of the models defined here, and an alternative dark matter model, which we will call model $Y$ (this could \eg be axions, primordial black holes, fuzzy dark matter, etc.). Note that in general, a Bayes' factors $B\sim 1$ shows no model preference, while a large/small value indicates preference for the model in the numerator/denominator.  

Now imagine further that no experiments to-date have tested model $Y$ -- that is to say, that the Bayesian evidence for model $Y$ is the same today as it was circa 1970. In this case, $P_{\rm current}$ is nothing more than the ratio of the Bayes' factor today to the value prior to WIMP experimental data, \ie
\begin{equation}
P_{\rm current} = \frac{B_{\rm current}}{B_{\rm pre-bounds}} = \frac{\int \, d\Theta \, P(X_{\rm current}|\Theta) \, P(\Theta)}{\int \, d\Theta \, P(X_{\rm pre-bounds}|\Theta) \, P(\Theta)} \, .
\end{equation}
Suppose that for a given model $B_{\rm pre-bounds} \sim 10$, and the value of $P_{\rm current}$ is found here to be $\sim 0.01$; in this case, one should conclude that a model which was once viewed as quite favorable relative to model $Y$, should today be thought of as less favorable. Clearly, the asymptotic behavior of $P_{\rm current} \rightarrow 1$ ($P_{\rm current} \rightarrow 0$) coincides with the desired limit that current experimental observations have not probed (or entirely probed) the model of interest.

The definition of \Eq{eq:pcur} can be easily generalized to determine the probability $P_{\rm future}$ of excluding a particular model by the time direct detection experiments reach the neutrino floor. These results are computed using a personalized code and tabulated in Tables~\ref{bayestable2} and~\ref{bayestable1}, for the case of Dirac and Majorana dark matter, respectively. In the case of $Z'$ mediated Dirac dark matter, the vast majority of the parameter space is already ruled out ($P_{\rm current} \lesssim 0.02$), regardless of which of these three priors we adopt. In the Majorana case, however, substantial portions of the parameter space remain viable, with $P_{\rm current}$ typically falling in the range of 10\% to 80\% in the case of log-flat priors. An exception to this are those Majorana models in which the $Z'$ couples significantly to first generation leptons, which are more significantly constrained. Future direct detection experiments are projected in most cases to explore between 20\% and 80\% of the currently viable parameter space, depending on the scenario considered and which priors are adopted.

\section{Discussion and Summary}
\label{conclusion}

Although WIMPs have long been viewed as among the most well-motivated classes of dark matter candidates, this paradigm has come to be seen as less attractive in the light of recent experimental constraints. In this study, we set out to explore and, to some degree, quantify the extent to which this reaction is warranted. To this end, we focused on the case of models in which the dark matter annihilates through a new gauge boson, $Z'$. While certainly not an exhaustive examination of all possible WIMP scenarios, this does provide us with a representative subset of models that we can use to consider the status of the WIMP paradigm.

In the course of this study, we have considered a wide range of scenarios featuring different charge assignments, masses, and couplings, as well as dark matter candidates that are either Dirac or Majorana fermions. We have then determined in each case the fraction of the initially viable parameter space that has been ruled out by existing experiments, as well as the fraction that is projected to be within the reach of future direct detection experiments (with sensitivity near the neutrino floor, as discussed in Sec.~\ref{neutrinofloor}). As these results depend on the Bayesian priors that we adopt on the parameter space, we consider three different sets of priors (see Sec.~\ref{bayesian}) and allow the reader to weigh them as they see fit.

Some of the main results of our analysis include:
\begin{itemize}
\item{In the case of Dirac dark matter, the vast majority of the parameter space is already ruled out by a combination of constraints from direct and indirect detection experiments, as well as observations of the cosmic microwave background  (the probabilities are $\lesssim 2\%$ for each choice of priors, as shown in Table~\ref{bayestable2}). The small regions that are not currently excluded are projected to be within the reach of upcoming direct detection experiments.}
\item{In the case of Majorana dark matter and a $Z'$ that is coupled to quarks, the current constraints are significant, but less restrictive. Across the range of charge assignments and coupling ratios considered, we find probabilities that fall between 4\% and 76\% (see Table~\ref{bayestable1}). These models are most significantly constrained by direct detection experiments, the LHC, and a series of lower energy accelerator experiments.}
\item{Scenarios featuring Majorana dark matter and a $Z'$ with substantial couplings to first generation leptons are strongly constrained. In particular, measurements from LEP, Borexino, and lower energy accelerator experiments strongly restrict this class of models.}
\item{We project that future direct detection experiments (with sensitivity near the neutrino floor) will in most cases be sensitive to between 20\% and 80\% of the currently viable parameter space, depending on which scenario is considered and the priors that are adopted. This provides significant motivation for the next generation of direct detection experiments.}
\item{Scenarios in which the $Z'$ couples more strongly to the dark matter than to SM particles are often much less stringently constrained, although future direct detection experiments will explore much of this parameter space.}
\end{itemize}

Throughout this study, we have defined WIMPs as stable particles that were in equilibrium in the early universe, and that annihilated into SM particles in order to yield a thermal relic abundance equal to the measured cosmological dark matter density. Across much of the parameter space, the determination of the thermal relic abundance depends on the product of the mediator's couplings to the dark matter and to the SM final states, $\Omega_{\chi} h^2 \propto (g_{\chi} \, g_{\rm SM})^{-2}$. In the case of $g_{\chi} \gg g_{\rm SM}$, however, the dark matter could annihilate directly into $Z' Z'$ (if kinematically allowed, $m_{\chi} \gtrsim m_{Z'}$), leading to a very different phenomenological picture. In particular, such hidden sector scenarios are less easily tested with direct detection experiments, or at the LHC and other accelerators~\cite{Pospelov:2007mp,ArkaniHamed:2008qn,Abdullah:2014lla,Berlin:2014pya,Martin:2014sxa,Hooper:2012cw,Berlin:2016vnh,Berlin:2016gtr,Dror:2016rxc,Dror:2017gjq,Feng:2009mn,Cheung:2010gj,Chu:2011be,Escudero:2016ksa,Zurek:2013wia}. While the null results of such experiments have provided additional motivation for this class of dark matter models, we do not consider them to lie within the boundaries of the WIMP paradigm and thus did not explore them in this study (limiting $g_{\chi} \le 10 \, g_{\rm SM}$).

\begin{acknowledgments} 
The authors thank Andrew Fowlie for pointing out some misleading language used in \Sec{bayesian}.  
CB is supported by the US National Science Foundation Graduate Research Fellowship under grants number DGE-1144082 and DGE-1746045. ME is supported by the European Research Council under the European Union's Horizon 2020 program (ERC Grant Agreement No 648680 DARKHORIZONS). This manuscript has been authored by Fermi Research Alliance, LLC under Contract No. DE-AC02-07CH11359 with the U.S. Department of Energy, Office of High Energy Physics. SW is supported by the Spanish grant FPA2017-85985-P of the MINECO, and by the European Union's Horizon 2020 research and innovation program under the Marie Sklodowska-Curie grant agreements No.~690575 and 674896.

\end{acknowledgments}

\bibliography{SimplifiedZPModels.bib}

\begin{appendix}
\section{Partial Wave Unitarity} \label{sec:pwu}

Following the arguments described in Ref~\cite{Griest:1989wd} (see also Ref.~\cite{Smirnov:2019ngs}), it is possible to use considerations involving partial wave unitary to derive a model-independent upper bound on the mass of the dark matter. While limited exceptions to these conclusion can be found in models in which the dark matter annihilates through a narrow resonance, for example, these constraints are quite general, and cover a wide range of dark matter candidates that are thermal relics of the early universe.

The thermal relic abundance of a species is given by the following:
\begin{eqnarray}
\label{unit1}
\Omega_{\chi} h^{2}& \simeq &\frac{(n+1) x_{f} \,1.07 \times 10^{9}\, \mathrm{GeV}^{-1}}{g_*^{1 / 2} m_{\mathrm{Pl}}\left\langle\sigma v\right\rangle_{f}} \\
 & \simeq & 0.12 \times (n+1) \bigg(\frac{x_{f}}{25}\bigg) \bigg(\frac{2 \times 10^{-26}  \text{cm}^3 / \text{s}}{\left\langle\sigma v\right\rangle_{f}}\bigg), \nonumber
\end{eqnarray} 
where, $g_*$ is the effective number of degrees of freedom at freeze-out, $m_{\rm Pl}=1.22\times10^{19}$ GeV is the Planck mass, $x_f \equiv m_{\chi}/T_{f} \approx 25$, $T_f$ is the temperature at freeze-out, and $n$ is defined such that $\left\langle\sigma v\right\rangle_{f} \propto v^n $. Expanding the cross section in terms of partial waves, $\sigma = \sum_{j} \sigma_j$, the requirement of unitary imposes the following constraint (for each $j$):
\begin{align}
\sigma_j v \leq \frac{4\pi(2j+1)}{m_{\chi}^2 v} \approx 6 \times 10^{-23}\sqrt{x_f} \; \text{cm}^3 / \text{sec} \left(\frac{m_{\chi}}{1 \, \text{TeV}}\right)^{-2}.
\label{unit2}
\end{align}
We see that $j=0$ imposes the strongest constraint. Combining Eq.~\eqref{unit1} and Eq.~\eqref{unit2}, we arrive at the following:
\begin{align}
m_{\chi} \leq 117\; \text{TeV} \left( \frac{\Omega_{\chi} h^{2}}{0.12}\right)^{1/2} \left( \frac{2}{g_{\rm dof}}\right)^{1/2} \left( \frac{25}{x_f}\right)^{1/4},
\label{unit3}
\end{align}
where $g_{\rm dof}=2 \, (4)$ for a thermal relic that is a Majorana (Dirac) fermion.


While the well-known constraint of Eq.~\eqref{unit3} is powerful and quite general, one can also apply arguments based on partial wave unitarity to specific models, deriving in some cases even more stringent constraints. More specifically, for a particular process with a scattering matrix element, $M(\theta)$, the requirement of partial wave unitarity states that 
\begin{align}
\label{A4}
\sum_{j=0}^{\infty} \left(2j+1\right) \text{Im}(a_{\mu \mu\prime}^j)\geq \frac{2\lvert\vec{p_i}\rvert}{E_{CM}} \sum_{j=0}^{\infty} \left(2j+1\right) \lvert a_{\mu \mu\prime}^j \rvert^2 ,
\end{align}
where $j$ is the total angular momentum quantum number, $\mu$ ($ \mu\prime $) are defined as $\mu = \frac{1}{2}(\lambda_2 - \lambda_1)$ and $\mu\prime = \frac{1}{2}(\lambda\prime_2 - \lambda\prime_1)$, given the spin of the initial (final) state fermions $\lambda$ ($\lambda\prime$)~\cite{Schwartz:2013pla}. The coefficients in the angular momentum expansion of the matrix element, $M_{\mu \mu\prime}(\theta)$, are given by
\begin{align}
\label{A5}
a_{\mu \mu\prime}^j = \frac{1}{32\pi} \int_{-1}^{1} d(\text{cos}(\theta)) d_{\mu \mu\prime}^j M_{\mu \mu\prime}(\theta),
\end{align}
where $d_{\mu \mu\prime}^j$ are the Wigner (small) $d$-matrices~\cite{Kahlhoefer:2015bea}. The first two matrices are $d_{0,0}^0$ = 1, and 
\begin{align}
d_{\mu,\mu\prime}^1&=\left(
\begin{array}{ccc}
 \cos ^2\left(\frac{\theta }{2}\right) & -\sqrt{2} \cos \left(\frac{\theta }{2}\right) \sin \left(\frac{\theta }{2}\right) & \sin ^2\left(\frac{\theta }{2}\right) \\
 \sqrt{2} \cos \left(\frac{\theta }{2}\right) \sin \left(\frac{\theta }{2}\right) & \cos (\theta ) & -\sqrt{2} \cos \left(\frac{\theta }{2}\right) \sin \left(\frac{\theta }{2}\right) \\
 \sin ^2\left(\frac{\theta }{2}\right) & \sqrt{2} \cos \left(\frac{\theta }{2}\right) \sin \left(\frac{\theta }{2}\right) & \cos ^2\left(\frac{\theta }{2}\right) \\
\end{array}
\right).
\end{align}
In the Majorana case and in the absence of a scalar, this can be used to set bounds on the strength of the couplings. Taking the $j=0$ term for dark matter self-scattering in the center-of-momentum frame, $\chi\chi\rightarrow\chi\chi$, Eq.~\eqref{A4} becomes
\begin{align}
\lvert\text{Re}(a_{00}^0)\rvert \leq \frac{1}{2v}.
\end{align}
To evaluate this expression, only the $\mu=\mu\prime=0$ amplitude is needed. The $s$, $t$, and $u$-channel diagrams contribute to the amplitude and we get the following expression:
\begin{align}
M_{0,0}(\theta)&= \frac{-8g_\chi^2 p^2 (1-\text{cos}(\theta)) \left(\frac{2m_\chi^2}{m_{Z^\prime}^2} -1\right)}{-p^2(1-\text{cos}(\theta))-m_{Z^\prime}^2 + i m_{Z^\prime}\Gamma} + \frac{8g_\chi^2 p^2 (1+\text{cos}(\theta)) \left(\frac{2m_\chi^2}{m_{Z^\prime}^2} -1\right)}{-p^2(1+\text{cos}(\theta))-m_{Z^\prime}^2 + i m_{Z^\prime}\Gamma}+\frac{16g_\chi^2 m_\chi^2\left(\frac{s}{m_{Z^\prime}^2}-1\right)}{s-m_{Z^\prime}^2 + i m_{Z^\prime}\Gamma}.
\end{align} 
Performing the integral in Eq.~\eqref{A5}, the $j=0$ coefficient is given by
\begin{align}
a^0 = \frac{-g_\chi^2 m_{\chi}^2 \left(m_{\chi}^2\left(4+v^2\right)-m_{Z^\prime}^2\right)}{m_{Z^\prime}^2 \pi \left(m_{Z^\prime}\left(m_{Z^\prime}-i\Gamma\right)-m_\chi^2 \left(4+v^2 \right)\right)}.
\end{align}
Finally, the unitarity condition is given by the following:
\begin{align}
g_{\chi }^2\leq \frac{\pi  m_{Z^\prime}^2}{2m_{\chi }^2 v}\left(1+\frac{\Gamma ^2 m_{Z^\prime}^2}{\left(m_{Z^\prime}^2-m_{\chi }^2 \left(v^2+4\right)\right)^2}\right).
\end{align}
Note that for $v \rightarrow 1$ and $\Gamma \rightarrow 0$, this bound converges to the result presented in Ref.~\cite{Kahlhoefer:2015bea}, $g_{\chi }^2\leq \pi  m_{Z^\prime}^2/2 m_{\chi }^2$. At freeze-out, however, $v \approx \sqrt{6/x_f}$ and $x_f \approx 25$, leading this bound to be relaxed by a factor of $\sim$\,$2$. In the Dirac case, we find that this calculation yields a constraint that is less stringent than that shown in Eq.~\eqref{unit3}.

When a scalar is present in the theory, the constraints on the couplings relax. Repeating the above calculation for $\rho \rho \rightarrow \rho \rho$ scattering, we get the following bound on the mass of the scalar~\cite{Kahlhoefer:2015bea}:
\begin{align}
m_{\rho}\leq \frac{\sqrt{\pi}  m_{Z^\prime}}{g_\chi }.
\end{align}
Throughout this work, we set the value of $m_\rho$ such that it saturates this bound, in order to minimize the phenomenological consequences of this particle.


\end{appendix}




\end{document}